# IGM Emission Observations with the Cosmic Web Imager: I. The Circum-QSO Medium of QSO 1549+19, and Evidence for a Filamentary Gas Inflow

D. Christopher Martin[1], Daphne Chang[1,2], Matt Matuszewski[1], Patrick Morrissey[1], Shahin Rahman[1], Anna Moore[3], Charles C. Steidel[4]

[1]Cahill Center for Astrophysics, California Institute of Technology, 1216 East California Boulevard, Mail Code 278-17, Pasadena, California 91125, USA. [2]Deceased. [3]Caltech Optical Observatories, Cahill Center for Astrophysics, California Institute of Technology, 1216 East California Boulevard, Mail Code 11-17, Pasadena, California 91125, USA. [4]Cahill Center for Astrophysics, California Institute of Technology, 1216 East California Boulevard, Mail Code 249-17, Pasadena, California 91125, USA

Running Title: Cosmic Web Imager Observations of Gas Near QSO 1549+19

**Abstract**

The Palomar Cosmic Web Imager (PCWI), an integral field spectrograph designed to detect and map low surface brightness emission, has obtained imaging spectroscopic maps of Lyα from the circum-QSO medium (CQM) of QSO HS1549+19 at redshift z=2.843. Extensive extended emission is detected from the CQM, consistent with fluorescent and pumped Lyα produced by the ionizing and Lyα continuum of the QSO. Many features present in PCWI spectral images match those detected in narrow-band images. Filamentary structures with narrow line profiles are detected in several cases as long as 250-400 kpc. One of these is centered at a velocity redshifted with respect to the systemic velocity, and displays a spatially collimated and kinematically cold line profile increasing in velocity width approaching the QSO. This suggests that the filament gas is infalling onto the QSO, perhaps in a cold accretion flow. Because of the strong ionizing flux, the neutral column density is low, typically $N(HI) \sim 10^{12} - 10^{15} cm^{-2}$, and the line center optical depth is also low (typically $\tau_0 < 10$), insufficient to display well-separated double peak emission characteristic of higher line optical depths. With a simple ionization and cloud model we can very roughly estimate the total gas mass ($\log M_{gas} = 12.5 \pm 0.5$) and the total ($\log M_{tot} = 13.3 \pm 0.5$). We can also calculate a kinematic mass from the total line profile ($2 \times 10^{13} M_\odot$), which agrees with the mass estimated from the gas emission. The intensity-binned spectrum of the CQM shows a progression in kinematic properties consistent with heirarchical structure formation.

## 1. Introduction

Theories of cosmological structure formation in a Cold Dark Matter (CDM) dominated universe predict that the matter in the intergalactic medium (IGM) and the galaxies formed from it are organized in a "cosmic web" of walls and filaments (Bond, Kofman, & Pogosyan 1994; Davis et al. 1985; Frenk et al. 1985; Miralda-Escude et al. 1996) – with structures spanning scales ranging from about one megaparsec to a few gigaparsecs (Mpc, Gpc, comoving). The intersections of sheets and filaments mark regions of high overdensity in which galaxies form and thrive, while the vast spaces between them are devoid of matter in comparison. Galaxy redshift surveys uncovered this morphology on large scales several decades ago (de Lapparent, Geller, & Huchra 1986). The gas associated with the filaments has been probed for nearly half a century via absorption features in the spectra of high redshift quasars (Gunn & Peterson 1965; Lynds 1971). A comprehensive review of the derived IGM properties is given in Meiksin (2009). These investigations have revealed the evolution of the IGM, its enrichment with metals and association with galaxies. For instance, the spectral density of the Lyman alpha (Ly$\alpha$) forest lines at redshift z ~ 3 indicates that nearly all baryons in the universe at that time were in a relatively cool phase (T ~ $10^4$K) (Rauch et al. 1997). At lower redshifts, near the current epoch, the Ly$\alpha$ forest thins out considerably with a large fraction of the cool gas being shock heated as it undergoes further gravitational collapse within the filaments, giving rise to a warm-hot intergalactic medium (WHIM) (Bregman 2007). Some of this gas accretes onto the virialized dark matter halos coalescing onto galaxies. Numerical simulations of structure formation are consistent with this picture, suggesting that cosmic web forming baryonic matter traces the underlying dark matter well (Cen et al. 1994; Cen & Ostriker 2006; Fukugita & Peebles 2004; Hernquist et al. 1996; Prochaska & Tumlinson 2009; Zhang, Anninos, & Norman 1995). The formation and evolution of galaxies is expected to be intimately connected to the IGM. Recent modeling indicates that galaxies may acquire a large fraction of their stellar fuel via streams of cold gas flowing along IGM filaments. This gas can fuel rapid periods of star formation and could explain the abundance of starburst galaxies observed around z ~ 2 (Birnboim & Dekel 2003; Faucher-Giguère, Kereš, & Ma 2011; Kereš et al. 2005; van de Voort et al. 2011a; van de Voort et al. 2011b). The computer models are being continually refined, however, resulting in modifications of the predictions of the relative impact of "cold" and "hot" mode gas accretion (Nelson et al. 2013). Observations of the morphology and kinematics of the IGM around galaxies and its correlation and co-evolution with them are necessary to vet the computer models and to further test our understanding of cosmic structure and galaxy formation and evolution (Faucher-Giguère & Kereš 2011; Fumagalli et al. 2011; Kirkman & Tytler 2008). While QSO absorption sightlines have yielded volumes of information about the co-evolution of IGM gas and galaxies, the information is limited in nature due to the sparseness of suitable background continuum sources on the sky and is, largely, statistical. Probing the full three-dimensional structure of specific systems requires another approach – either relying on a different population of background sources (more numerous or extended; nominally galaxies) or measuring emission from the systems of interest. The former is likely not feasible until the 30 meter class telescopes come online, while the latter is within the capabilities of current instrumentation, particularly in the case of systems developing in regions of high overdensity.

Emission in strong UV resonance lines, particularly HI Lyα1216Å, is predicted to delineate the cosmic web (Binette et al. 1993; Cantalupo et al. 2005; Furlanetto et al. 2003, 2005; Gould & Weinberg 1996; Hogan & Weymann 1987; Kollmeier et al. 2010). For the vast majority of gas within the cosmic web the emission is expected to be faint (~$10^{-20}$ erg/s/cm$^2$/arcsec$^2$), as the only source of ionizing radiation is the relatively weak metagalactic UV background. Lyα fluorescence from optically thick IGM illuminated by strong ionizing sources, typically QSOs, is thought to be boosted above the typical level by several orders of magnitude (Cantalupo, et al. 2005). Another energy source anticipated to fuel detectable emission that needs to be taken into account when characterizing circum-galactic environments is collisional excitation due to gravitational infall of gas, for example from cold accretion flows (Fardal et al. 2001; Faucher-Giguère et al. 2010; Förster Schreiber et al. 2006; Haiman & Rees 2001). The expected surface brightness from the infalling gas is estimated to be on par with Lyα fluorescence due to QSO ionizing radiation (Dijkstra, Stuart, & Wyithe 2006).

Significant effort is being exerted to detect and characterize the IGM in emission. Deep narrowband observations (Fynbo, Moller, & Warren 1999; Hayashino et al. 2004; Nilsson et al. 2006; Steidel et al. 2000) have imaged numerous Lyα blobs in overdense regions while Cantalupo, Lilly, and Haehnelt (2012) report a detection of a rich field of Lyα sources around a z~2.4 quasar. Various techniques, both imaging and spectroscopic, have been used to detect Lyα halos around galaxies and quasars (Francis & Bland-Hawthorn 2004; Hennawi et al. 2009; Hu & Cowie 1987; North et al. 2012; Steidel et al. 2011). Long- and multi-slit spectroscopic investigations have resulted in detections of a number of Lyα emitters in the vicinity of quasars (Rauch et al. 2008). More recently, spectroscopic surveys have yielded intriguing observations of Lyα emission from the IGM around quasars and galaxies exhibiting possible filamentary nature and velocity structure (Hennawi & Prochaska 2013; Rauch et al. 2013). There have also been reports of emission associated with damped Lyα systems near proximate QSOs, although the observed flux is thought to originate from either the DLA host galaxy or the background quasar (Adelberger et al. 2006; Hennawi, et al. 2009). The evidence appears strong that volumes around galaxies and quasars are rich in gas that manifests a complex morphology, kinematic structure, and interacts extensively with the central object. Unfortunately, the above-outlined observations and techniques do not result in a full picture of the individual systems being studied, yielding either incomplete spatial or limited spectral information. Integral field spectroscopy, however, combines moderate resolution spectral content with spatial data over a compact field that is well matched in shape and size to the circum-galactic and circum-QSO environments. As such, it is a great tool for characterizing galaxies and their neighborhoods. We have initiated a program to study faint emission around galaxies, quasars, and Lyα nebulae with the Palomar Cosmic Web Imager, an integral field spectrograph for the Hale telescope at Palomar Observatory built specifically to detect and map dim and diffuse emission.

In this paper we report the on the detection and characterization of extended Circum-QSO Medium (CQM) gas surrounding QSO HS1549+1919 with PCWI. In §2 we describe the instrument, data and analysis. In §3 we compare the PCWI results to the narrow-band image of the QSO. In §4 we describe our smoothing algorithm (3D Adaptive Smoothing including Lambda [3DASL]), which we use to display image slices of the observed spectral data cube. Section §4.2 is a detailed technical discussion of robustness tests for this algorithm and the results. We show a number of simulations and additional information which we hope builds confidence in the applicability and fidelity of this algorithm to moderate signal-to-noise spectral image cubes. This section can be skipped without loss of continuity. In §5 we study the morphology and derive physical parameters from the data and simple models. We summarize the conclusions in §6. We use a standard WMAP-7 cosmology (Komatsu et al. 2011), and unless otherwise specified all distances are given in proper (physical) coordinates.

## 2. Cosmic Web Imager Data and Analysis

### 2.1. The Palomar Cosmic Web Imager

We have constructed an integral field spectrograph called the Palomar Cosmic Web Imager (PCWI), that is designed to search for, map, and characterize IGM emission and other low surface brightness phenomena (Matuszewski et al. 2010). It uses a 40" x 60" reflective image slicer with 24 40" x 2.5" slices. For these observations the spectrograph and slicer was oriented with individual slices running east-west. The spectrograph has a high dispersion volume phase holographic grating covering 4500-5400Å with an instantaneous bandwidth of 400Å. The spectrograph attains a slit-limited spectral resolution of $\Delta\lambda \sim 1$Å and a peak efficiency of ~10% at 5000Å including the telescope and atmosphere. PCWI is mounted at the Cassegrain focus of the Hale 5 meter telescope on Mt. Palomar. As a spectroscopic imager it is possible to form individual 1Å wide images of the sky. The imaging resolution, while limited by the 2.5" slicer sampling, can be effectively improved to ~1.3" by dithering the field between individual exposures.

### 2.2. Observations of QSO HS1549+19

Since it is likely that IGM emission is brightest in regions near bright QSOs, we observed the region around QSO HS1549+1919 (z=2.843) that is known to harbor extensive narrow-band Lyα emission (Trainor & Steidel 2012). Figure 1 shows a narrow-band Lyα image obtained at Keck.

We obtained a total of 6.5 hours on-source and 6.5 hours off-source exposure centered on QSO HS1549+19 ($15^h51^m52.48^s$, +19°11´4.2´´), over a total region 120 arcsec East-West by 60 arcsec North-South. Individual exposures were obtained using a nod-and-shuffle technique developed for long-slit observations (Sembach & Tonry 1996). We used the central 1/3 of the CCD for recording the spectrum, and masked the outer 2/3 of the CCD (restricting the spectral bandpass to ~150Å). Each frame was created using the following proceedure. The telescope is pointed at the source region, the guide camera locks onto a guide star, the shutter is opened, and a two-minute exposure is taken. Then the shutter is closed, the telescope nodded to a background location 1.5 arcminutes north and 2 arcminutes west, the CCD is clocked by 670 pixels up, the shutter is opened, and a background image is obtained. In order to make an unbiased background subtraction the precise sequence is 1 minute background, 2 minutes source, 2 minutes background, …, 2 minutes source, 1 minute background for a total of 11 background and 10 source exposures, each a total of 20 minutes. A total of 10 of these source/background cycles was performed (a total of a little more than 40 minutes with overhead), followed by CCD frame readout. A total of 14 20/20 minute source/background and 11 10/10 minute exposures were obtained over 6 observing nights in 2011 and 2012. The objective of this method is to provide the highest precision sky subtraction possible without increasing the read noise. The CCD was read-out in a 2 by 2 binning mode to reduce the impact read noise further. The CCD read noise was less than 3 $e^-$ (rms).

The PCWI field of view is 40 arcsec (parallel to slices) by 60 arcsec (sum of 24 2.5 arcsec wide slices). This observation was generated from a mosaic of multiple pointings covering a total region of 120 arcsec by 60 arcsec. During this observation PCWI was oriented with slices in the east-west direction. In some regions of the mosaic several north-south pointings were dithered by fractional slit widths to provide the advantage of further averaging over instrument features and slice-to-slice efficiency variations. Background pointings were dithered typically by 0.5-2 arcsec. The background area was chosen because it has minimal bright narrow-band or continuum sources. We also obtained numerous calibration images using lamps throughout the observing period.

### 2.3. PCWI Pipeline Analysis

The PCWI pipeline converts CCD images into a 3D data cube. All individual exposures are bias-subtracted using a superbias, and cosmic rays are removed. Detector images are then split into source and background regions. Continuum flats are used to define the slice regions on the detector. An edge calibration image using a continuum lamp corrects is used to refine the horizontal registration of each CCD row, and a pinhole image is used to fix the plate scale in each slice. An arc lamp spectrum is sliced and the line tilts measured to correct for spectral line curvature in each slice. The resulting dewarped spectrum is then calibrated against the standard input spectrum to yield a polynomial slice-by-slice wavelength solution.

Science images are sliced, edge rectified, rescaled, dewarped, and wavelength rectified into source and background slice spectra. They are wavelength-shifted to compensate for <1Å of total flexure using sky lines. The final mosaiced data cubes are generated using the astrometry based on the guide star camera, and astrometry was checked using several moderately bright objects in the source fields. Astrometry is good to about 0.5 arcsec (rms). Mosaiced data cubes were created for source and background. Source data cubes are rectified based on astrometry, and background data cubes were created using the same source data cube registration (so that individual source and background exposures are subtracted one-to-one). Exposure maps are generated by first generating a smoothed "skyflat" from background spectral images in detector coordinates. These are normalized, exposure-time multipled, and mosaiced in the same way as source and background images. The result is a set of flux data cubes (RA, DEC, and λ) for each exposure, sampled at (0.292", 0.292", either 0.25Å or rebinned to 1Å) covering 4600-4750Å. The difference cube is obtained by simply subtracting source and background flux cube. Because the nod-and-shuffle mask does not physically contact the CCD, a small amount of diffuse continuum light remains in the subtracted cube (<1%), which is easily removed with a low-order continuum fit.

In the resulting difference cube, structures spanning 10" x 10" x 4Å have a 1-sigma noise level of ~1400 ph $cm^{-2}$ $s^{-1}$ $sr^{-1}$ $Å^{-1}$ (Line Units/Å = LU/A). Sky levels were ~1 x $10^5$ LU/Å (ph $cm^{-2}$ $s^{-1}$ $sr^{-1}$ $Å^{-1}$). Our emission line detections are at the level of ~3-10% of sky. Sky levels typically varied smoothly and mostly linearly by 0.5-2% on 20 minute timescales or typicaly 0.1% on the 2 minute nod-and-shuffle cadence, leading to sky-subtraction errors significantly less than 300 LU/Å for individual exposures, and correspondingly smaller errors for the summed exposures. 1000LU corresponds to ~$10^{-19}$ erg $cm^{-2}$ $s^{-1}$ $arcsec^{-2}$ at 5000 Å.

### 3. Comparison of PCWI data to the Narrow-Band Image

We would like to understand to what extent features detected in the PCWI data are present in and confirmed by the narrow-band image. In Figure 1 we show the narrow-band image obtained at Keck, along with broad and narrow slice images obtained with PCWI. The narrow-band image is centered on 4670Å with an 80Å width. Extensive emission is seen near the QSO, generally of a clumpy appearance. We have outlined several of these clumps in Figure 1a and in the other figure panels. In Figure 1b we show a broad (32Å) slice sum obtained from PCWI. The sum is produced using the smoothed data cube described in the next section. Clumps corresponding to the three most prominent clumps in the narrow-band image are observed in PCWI as well, albeit at lower spatial resolution and therefore with less distinction. In Figure 1c-d we show narrower PCWI slices centered on two prominent emission features which we later label as filament 1 (easterly extension centered at 4682Å) and filament complex 2 (westerly network of extended emission centered at 4672Å). In narrow slices the emission becomes more extended and filamentary and of lower intensity. In both cases the clumps are at the base (with respect to the QSO) of the more extended, narrow spectral width extensions. As we show later the clumps are much broader than the extended filaments. The narrow-band image is better at detecting spectrally broad emission clumps while PCWI is better at detecting extended, spectrally narrow emission features. Spectral confusion and to a lesser extent the increased noise in the broader band narrow-band image makes it less ideal for detecting narrow, filamentary features, while the clumps appear less distinct in the PCWI data because of lower resolution and greater noise in the wide bin slice. Because of the detection of extensive, narrow emission in the eastern third of the mosiac near the systemic wavelength (filament complex 2), we have added additional regions in order to make a quantitative comparison with the narrow-band image (shown in Figure 1a and 1b) in order to test for a spurious origin for filament complex 2.

In Figure 2 we show a comparison of the emission in these regions between the PCWI and narrow-band fluxes. The narrow-band image has a nominal Poisson noise level of $2.5 \times 10^{-19}$ erg cm$^{-2}$ s$^{-1}$ arcsec$^{-2}$ (2500LU) with a 1 arcsec$^2$ smoothing box. Because of the narrow-band filter width (80Å), typical sky (50,000LU/Å) corresponds to a total level of $4 \times 10^6$ LU. A 0.1% error in continuum subtraction would lead to a 4000LU error in the corrected narrow-band image. Continuum-subtraction errors in the narrow-band images are introduced by a number of factors, including the spatial/spectral variation in the instrument response over the narrow vs. broad continuum filter, the variation in the sky spectrum over the two wavelength ranges, the impact of continuum source wings and scattered light, and the additional narrow-band objects present in the broader continuum image. Spatially extended structures are particularly subject to these errors. We estimate from variations in the zero level (which occasionally goes negative) that typical systematic continuum-subtraction errors are at least ~0.1-0.2% of sky, or ~4-8 $\times 10^{-19}$ erg

cm$^{-2}$ s$^{-1}$ arcsec$^{-2}$ (~4000-8000LU), depend on the spatial scale, and may be higher in some regions. There are large, low spatial frequency variations that lead to the flux offset of ~10 x 10$^{-19}$ erg cm$^{-2}$ s$^{-1}$ arcsec$^{-2}$ (~10,000LU) apparent in Figure 2. The CWI image summed in ~5 x 5 arcsec$^2$ boxes (Figure 1b) has a similar Poisson error level (~10000LU). Thus for both images typical noise amplitude will cause variations from blue to green color values on 5 arcsec scales, as is observed. Comparison of individual features detected in either image will not be one-to-one because of this noise level. We note that the PCWI image has not had continuum sources removed, while the narrow-band image is continuum-subtracted. Given the uncertainties, the fluxes measured in the various regions are consistent within statistical and continuum subtraction errors.

To confirm the spectral content of these regions and the fidelity of the smoothing algorthm in reproducing this spectral/spatial content, we show the spectra obtained in the blob regions outlined in Figure 1 in Figure 3 using the unsmoothed data cubes. Region *a* shows a narrow component at 4683Å, consistent with the redshift derived from Keck/LRIS follow-up observations of the clump. Region *b* shows a broad component centered on 4673Å, matching the redshift derived from Keck follow-up on this clump. Region *c* shows a moderate width feature centered on 4670Å, again consistent with the redshift derived from Keck/LRIS follow-up. Finally, region *e*, which does not appear as a distinct emission component in the narrow-band image, but does appear as a filamentary connection between clump *b* and *c* in the PCWI 4672Å slice, shows a narrow component at 4671Å. This corresponds to the systemic velocity of the QSO. All of the observations are consistent with a trend analyzed in § 5.3 in which faint emission is spectrally narrow and more extended, while spectrally broad emission is associated with the brighter emission clumps. The total region around the QSO is displayed in the final panel, and a summary of fluxes, luminosities, and detection significance of the detected emission features is provided in Table 1. The emission is detected with high significance in all the regions.

We conclude that, taking into account the nature of the underlying emission and the relative strengths of the two observational techniques in detecting spatially extended, spectrally narrow emission, and spatially compact, spectrally broad emission, that the two observations are consistent.

## 4. Adaptive Smoothing Algorithm

In general, even long exposure data from PCWI targeting IGM emission is of modest signal-to-noise ratio. IGM emission models predict that line emission will be extended, probably filamentary, and may show significant kinematic variations with spatial position (Cantalupo, et al. 2005; Kollmeier, et al. 2010). We have therefore developed and extensively tested an adaptive smoothing algorithm designed to highlight extended line emission while not artificially smoothing more compact emission line and continuum sources.

### 4.1 Results of Adaptive Smoothing Algorithm

The slice images displayed in **Figure 4** (and following) are produced from the raw difference flux cube by a 3D adaptive smoothing algorithm (3DASL). The algorithm works as follows. The (0.292" x 0.292" x 1Å spaxel) data cubes are first boxcar smoothed spectrally by 2Å, then smoothed spatially (boxcar or Gaussian kernels), by successively larger smoothing kernels ranging from FWHM=3 pixels (1 arcsec) to 70 pixels (20 arcsec) in 2 pixel increments. Boxcar or Gaussian kernels produce similar results, but the latter exhibit less high-spatial-frequency noise. The effective signal-to-noise threshold is correspondingly reduced by a factor equal to the one pixel noise divided by the 1D smoothing width (e.g., a 1-sigma 1-pixel noise of 9000LU becomes 1800LU for a 5 pixel smoothing box, and 180LU for a 50 pixel box). Whenever a spaxel exceeds the noise threshold (which of course depends on the kernel) it is placed in the final cube (detected). The flux in the smoothed spaxel is subtracted from the unsmoothed cube pixel prior to the next spatial smoothing cycle. Any residual flux in the spaxel can still contribute to the smoothed slice images but is not available for further "detection". The unsmoothed difference cube is then smoothed in wavelength by a 4Å kernel, and the spatial smoothing algorithm is repeated. This loop repeats once more with a spectral smoothing of 8Å, or a total of three times. The 2.5-sigma noise threshold was derived directly from the difference cube, and is consistent with the predicted Poisson noise. We discuss below variations in the algorithm and signal-to-noise thresholds.

We show in **Figure 4** spectral images in 4Å slices obtained with this algorithm. We show in Figure 5 images that illustrate the raw data and noise in comparison to the adaptively smoothed image slices. We show later that pure noise produces a very different smoothed image and emission morphology. Many of the features apparent in the spectral images appear in multiple 1Å slices, as we show in Figure 6 and Figure 7.

## 4.2 Robustness Tests of the Adaptive Smoothing Algorithm

It is important to note that the principle role of the smoothed images is to identify extended emission in order to define regions that can be summed and studied, typically with the unsmoothed difference cubes. Because of the modest signal-to-noise ratio of the emission there will not be a detailed one-to-one correspondance of the smoothed slice images and the underlying data, particularly on small scales. As we show later with simulations, the algorithm tends to produce less extended and more clumpy emission than is underlying, again due to modest signal-to-noise. However as we demonstrate below there is a good correspondance on the scales of interest for this investigation.

**1. Simple filament simulations.** We have performed simulations to investigate the behavior of the algorithm under various conditions. We first show a set of simulations that are performed for a 300 x 300 pixel image (single wavelength slice). Figure 8 shows the first set of simulations which consist of a set of compact sources distributed in a "filamentary" pattern. 2DASL does a reasonable job isolating the sources, while simple smoothing tends to produce an artificial extended filament. In Figure 9 we add a long, narrow emission filament. The algorithm clearly revels extended, contiguous emission, although noise introduces small-scale structure in the filament which is not real. In Figure 10 we complicate the filament by adding a narrow core and broad wings. Again, 2DASL does a reasonable job revealing this more complex extended structure, while isolating the compact sources. In all three cases a mean of 10 realizations is quite close to the original image, showing that the while each individual image is still noisy, the mean of many images converges to the input image.

**2. Modifications to the signal-to-noise threshold and addition of bright central source.** For a pixel to be detected it must exceed the signal-to-noise threshold. The choice of the signal-to-noise threshold has a signficant effect on the resulting smoothed images. Our extended emission detections are modest in signal-to-noise ratio, so we have investigated the impact of this threshold on the resulting smoothed images. We show in Figure 10 and Figure 11 the impact of the threshold on the 2D image simulations. The presence of a very bright central source could impact the algortim, so we show in Figure 12 a simulation with such an added source. In this case the central source is not removed prior to smoothing, in order to check for any impacts. There is none outside a central zone a few arcseconds in radiuss.

In Figure **13** we have simulated a series of filaments with increasing intensity. This illustrates a detection threshold corresponding to $(S/N)*\sigma/\theta$, where $(S/N)$ is the smoothing algorithm threshold, $\sigma$ is the noise level (per pixel), and $\theta$ is the minimum feature size (narrow width of filament). In Figure 14 we show how the 3DASL results for real data in 4Å slices varies with S/N threshold from S/N>3.0 to S/N>2.0. Lowering the threshold has the principle effect of increasing the number of compact noise spikes detected by the algorithm, and very slowly increasing the size of extended features (real and noise). Based on this study we conclude that a conservative compromise is a threshold of S/N>2.5.

**3. Filament simulation including realistic noise, line profiles and velocity variations.** PCWI spectral images do not produce square image pixels, and are the result of many dithered individual exposures. This is because the coarse slice width (2.5 arcsec) undersamples the seeing PSF, and a single detector pixel is remapped to a 2.5 arcsec by 0.6 by 0.25Å arcsec spaxel in the data cube. We performed simulations of a data cube using the measured sky spectrum and spectral/spatial distortion to produce an input image (including source and background detector regions) with Poisson noise. Data cubes were generated corresponding to each QSO HS1549+19 exposure time and position using the standard pipeline. The individual exposure cubes were coadded in a mosaic and the source and background cubes differenced, following the identical steps as the actual data. The resulting raw, smoothed, and 3DASL smoothed simulated image slices are displayed in Figure 15 and compared to actual data. This figure shows that most of the noise is manifested as spurious compact emission regions, and only very occasionally is a low-level extended feature generated by random, contiguous regions that have spatial correlation partially produced by the tall, narrow spaxels. Comparing these images to **Figure 4** shows that the detected extended features are not a result of spurious smoothing of noise introduced by the 3DASL algorithm.

We add a filament with a flux level comparable to that detected in QSO HS1549+19 environment. We first assume the emission line has a width of 3Å and is fixed in velocity. The image slice at the line wavelength is given in Figure 16. The filament is clearly detected in the 4Å image slice. We show a "spectral-image" plot, in which we have sliced the smoothed data cube parallel to the filament axis and the wavelength direction, in Figure 17. The spectral image is summed over a "pseudo slit" of width 5 arcsec. This is a method to show how individual structures that may show kinematic variations along their length, which would render them as disconnected structurs in distinct image slices, may form a single extended feature. Simulations suggest that velocity variations could be hundreds of km/s along single filaments (Cantalupo, et al. 2005), while a single 1Å slice is 64 km/s wide. Finally, we added a velocity shear to the filament similar to that predicted in cosmological simulations. The resulting spectral-image plot also in Figure 17 shows that this shear can be detected and does not affect the detectability of the filament.

We also studied the effects of the dither pattern in each area of the mosaic. It can be seen from Figure 5 that the western third of the image mosaic has lower vertical resolution, comparable to the slice width 2.5 arcsec. Because of an inadvertant axes swap, these images were obtained without a half-slice north-south dither (perpendicular to the slices) used in the other 2/3 of the cube. The impact of this on the effective point spread function is to change the north-south 1D PSF from approximately a triangle function with FWHM = 2.5 arcsec to a square function with full width 2.5 arcsec. There should be no significant impact on the detectability of individual features broader than the slice width in the north-south direction, and very little additional susceptability to systemic error producing false detections than in the areas with a 1.25 arcsec north-south dither, since source and background data cubes are obtained in the identical fashion, and the dither is limited to 1.5 slices and therefore does not significantly reduce artifacts produced by an improperly flat-fielded single slice. We demonstrated this by generating a simulated filament extending to the west with intensity comparable to that detected in the actual data (corresponding to filament 2). The slice image and the spectral image are displayed in Figure 18.

**4. Features are detected in multiple slices.** We note that features detected in the data are clearly detected in multiple image slices. We identify two filaments in **Figure 4**, as discussed below in §5.1. These filaments are denoted filament 1 and filament complex 2. Filament complex 2 (centered on 4673Å) is detected in at least four 1Å slices (Figure 6). Filament 1 (centered on 4682Å) is detected in at least seven 1Å slices (Figure 7). While we expect some slice to slice correlation due to the wavelength component of the smoothing algorithm, pure noise does not produce such correspondance over many slices.

**5. Compact sources--impact on results.** A large number of faint compact sources could in principle be converted by the smoothing algorithm into apparent extended emission in the smoothed images. Such sources if unsubtracted could also add spurious emission to the unsmoothed data cube and spectra derived from regions. We have therefore carefully examined the impact of compact sources on the results.

We showed in Figure 8-10 that compact sources resolved by the instrument are also resolved by the smoothing algorithm, and in contrast to simple smoothing do not, as a rule, form spurious extended features. Of course a high surface density of compact sources would not be resolved by the instrument, and we cannot rule out that extended features are formed by the summation of many individual clumps or sources of Lyα emission. The narrowband image (Figure 1) suggests that there are emission-line clumps but that underlying extended components could also be present. Both of filaments 1 and 2 seem to be associated with clumps in the narrowband image that also are present in the PCWI slices. The clump associated with filament 1 appears in panel *h* of **Figure 4**, at 4984Å. The clump associated with filament complex 2 appears in panels *c* and *d* (4664 and 4968Å).

Continuum and line sources can be removed from the data cube to test their impact. We used a continuum image to determine the location and approximate brightness of all compact continuum sources in the field of view of our source field. We determined the point spread function (PSF) from prior observations. In this PSF 80% of the light is confined to a region of 7.4 arcsec$^2$. We used a source removal algorithm that uses this PSF and neighboring sky, slice by slice, to remove each source and replace it with an estimate of local sky (which could include local extended emission). There are a total of 84 sources detected in the continuum image of the source field to V<25. Of these, 18 are Lyα sources in the narrow-band image, which we did not remove. The remaining 66 continuum sources are removed. Figure 19 shows the compact source distribution and magnitudes superimposed on the narrow-band image, along with the boxes used for source definition. There are no significant differences in the resulting smoothed images or region spectra with or without source subtraction. This figure also shows an image comparison with and without sources subtracted. All images and spectra are given without sources subtracted.

## 5. Properties of Extended Emission

### 5.1. Morphological Properties

We begin by describing the distribution of emission as displayed in the image slice plots of **Figure 4**, Figure 6, and Figure 7. The emission is most prominent in the band 4660-4685Å. In slice 4664Å a large exension is present toward the north-west. Moving from 4664A to 4672Å the extension moves southward and grows to fill the western portion of the field. The cloud at $\Delta\alpha=-30$, $\Delta\delta=-7$, prominent in Figure 1, is clearly present in slice 4672Å. Clumps nearer to the QSO on the west side in Figure 1 are also visible here. The cloud and the QSO are bridged by emission along the east-west direction, as well as emission looping southward. In slice 4676Å the emission extends mostly to the south. In slice 4680Å a narrow filament extends to the east and slightly south, as far as 30 arscec from the QSO. In 4684Å the filament is more of a clump, a feature also present in Figure 1. Finally in 4688Å the extended emission has mostly vanished.

We described in the previous section how spectral-image plots can delineate what could be contiguous extended features exhibiting kinematic shear. We have generated an array of these plots in pseudo-slits centered on the QSO with varying position angle, shown in Figure 20. The extensive extended emission near the systemic velocity and apparent in **Figure 4** is also present in these plots. While extended emission is seen at all azimuths, regions show strong, narrow emission lines. Azimuth 345° shows a strong (#1, cf. below) near 4682Å. Azimuth 180°-210° (shown as $\Delta\theta<0$ for azimuth 0°-30°) shows extension out to r>50 arcsec near the systemic velocity.

In Figure 21 and Figure 22 we use the image slice to trace out pseudo-slits guided by the image morphology. Figure 21 shows filament 1 extends to at least r=30 arcsec, and may extend to greater than 40-50 arcsec. Double peaked emission could be present between 20 and 40 arcsec. The feature shows a central velocity near 4682Å, with modest velocity shear along its length, until r=12 arcsec. At this point the velocity shifts sharply to approach the QSO systemic value at 4672Å. This point also corresponds to a large clump in the narrowband image.

The emission complex appearing mainly around slice 4672Å can be described as extensive, compared to that in most of the other slices, and perhaps "web-like", with clumps embedded in a network of apparent filaments. We have delineated one path through these filament-like connections in Figure 22, which we denote filament 2. The extension in Figure 22 bridges the QSO with the large region of clumpy emission in the west. The emission is narrow and near the systemic velocity for r>27 arcsec. In the zone 20<r<30 arcsec the emission profile broadens sharply and is accompanied by continuum, at the location of a prominent line and continuum source. While we showed that that narrow-band image and CWI data are roughly consistent, because this complex does not have a clear counterpart in the narrow-band image, we regard this detection as more tentative than filament 1.

### 5.2. Physical Properties

We use a simple CLOUDY model (Ferland 1996) to derive physical properties of the CQM. We use the known QSO HS 1549+19 brightness, normalized at Ly$\alpha$, of $L_Q = 1 \times 10^{48}$ erg/s, and a standard assumed spectral energy distribution (AGN T=1.5 x $10^5$K, a(ox)=-1.4, a(uv)=-0.5, a(x)=-1) to derive the ionizing and Ly$\alpha$ photon-pumping continuum. The Ly$\alpha$ line is extremely broad and the continuum normalization includes it. For the reference model, we assume isolated clouds with a constant physical thickness (along the radial direction to the QSO) equal to 40 kpc at all radial distances (corresponding to 5 arcsec). The cloud distance from the QSO is variable. The clouds are assumed to have a constant density that is a free parameter, unit filling factor, and metallicity 10% solar to approximate abundances in circum-QSO gas. Dust absorption is ignored. The total intensity in Ly$\alpha$ is calculated for each radial cloud location, in a +/-20Å window around the systemic Ly$\alpha$ wavelength of 4672Å. The radial location and total intensity then implies a cloud density, column densities of N(HI) and N(H), mean gas temperature, etc.

We show the radial dependence of cloud density and intensity, compared with observed values, in Figure 24. Three processes produce Lya photons: radiative recombination, photon pumping, and collisional excitation. Because of the hyperluminous QSO, the gas is highly ionized and hot even when relatively dense, neutral hydrogen column densities are low near the QSO, and most Lyman continuum photons escape. The temperature run with radius is shown in Figure 26. The temperature is far higher than that in the general IGM and Lyman alpha forest. We investigate below whether the derived temperature impacts the conclusions. Based on the CLOUDY model output, collisional excitation accounts for at most 1-10% of the total line intensity, in spite of the high temperature, because the neutral fraction is very low. In most regions the Lyα line emission is dominated by *photon pumping*, or scattering of QSO continuum light in the Lyα line of the surrounding gas. In Figure 24 solid lines give intensity vs. radius for pumping and recombination (also called Lyα fluorescence or LAF). Dotted lines give the pumping contribution. Lines that are pumped will not typically show a double peaked profile, which arises when recombination photons are escaping a large optical depth in the line through the wings. The vertical line shows the minimum inner radius of the analysis, corresponding to 10 arcsec (78 kpc). All masses are calculated using emission only outside this radius. Inside this radius the wings from the QSO are too strong to allow secure conclusions. Note that we have also not included any gas cooling due to dissipation of gravitational energy as it accretes into the dark matter halo. This approximation is accurate for the total emission in the field of view, which has a flux of $1.7 \times 10^{-14}$ erg s$^{-1}$ cm$^{-2}$ (cf. Table 1) about 500 times higher than the flux predicted for a $7 \times 10^{12} M_\odot$ halo (Haiman, Spaans, & Quataert 2000).

From this reference model analysis we can therefore derive an approximate gas masses in the CQM. For filament 1 we derive a baryon (gas) mass of $M_{gas} \approx 1.2 \times 10^{11} M_\odot$. For the tentatively detected filament 2 we derive a baryon (gas) mass of $M_{gas} \approx 3.6 \times 10^{11} M_\odot$. In each case the mass is calculated by summing the data cube over a range $\lambda_{SYS} - \Delta\lambda < \lambda < \lambda_{SYS} + \Delta\lambda$, where $\Delta\lambda$ is 5-10Å for filaments and 20Å for the total field. We then convert the total intensity to a mass density using the model, and summing over either the filament or the full data cube not including the central 10 arcsec radius region around the QSO (area 3700 arcsec$^2$). The total Lyα luminosity of the observed data is $L_{Ly\alpha} \approx 1.3 \times 10^{46}$ erg/s, or 1.3% of the total QSO luminosity.

Using our simple model we derive a rough estimate of the total baryon (gas) mass of $M_{gas} \approx 2.9 \times 10^{12} M_\odot$. This implies a minimum total mass (gas plus dark matter) of $M_{tot} \approx 2.3 \times 10^{13} M_\odot$. This mass estimate is not changed significantly when we use the source subtracted data cube. At z=2.843, this total halo mass corresponds to a virial radius of $r_{vir} = 230$ kpc (or 30 arcsec) and circular velocity of $v_c$=650 km/s. As we see below, this is probably consistent with the average gas kinematics. We derived this total mass using the S/N>2.0 data cube. The corresponding values for thresholds S/N>2.5 and 3.0 are $M_{gas} \approx 2.5 \times 10^{12} M_\odot$ and $M_{gas} \approx 2.2 \times 10^{12} M_\odot$, indicating that less than 25% of the signal arises from 2.0<S/N<3.0 emission.

The total gas (baryon) mass of filament 1 and 2 is $M_{1+2} \approx 4.8 \times 10^{11} M_\odot$, while the total CQM gas (baryon) mass estimated above is $M_{1+2} \approx 3 \times 10^{12} M_\odot$, a factor of 6 higher. The definition of filament 2 is somewhat arbitrary since it is surrounded by an extensive emission complex. This filament 2 complex represents a significant contribution to the total (~50%). Other emission including a significant extension toward the north-north-west constitute the balance (see **Figure 4**). The bulk of the emission is very significantly detected (76% has S/N>3), and as we display in Figure 31 shows spectral characteristics consistent with velocity broadened Lyα emission. These two estimates conservatively bracket the range of possible gas masses for the CQM.

We now consider the uncertainties in the derived gas masses introduced by the model assumptions. Clearly, the radiative transfer of the optically thick Lyα line leads to signficant uncertainty and any kinematic trends which impact the line optical depth. However, because the ionizing field from the QSO is so strong, the HI column density is moderate near the QSO. In addition, the temperature of the gas is high. The combination of these two effects leads to moderate line center optical depths ($0.01 \leq \tau_0 \leq 100$), and in most regions $\tau_0 < 10$, as we show in Figure 25. Thus the number of scatterings is limited as is any absorption by dust in the CQM gas. Other than dust absorption, radiative transfer effects will mostly redistribute the line photons into the line wings and somewhat blur the correspondence between flux and local density (Kollmeier, et al. 2010). But at these low optical depths the effects are likely to be modest. A careful treatment of the effects of radiative transfer requires a more detailed geometrical and kinematic model of the CQM, and will be addressed in a future paper.

The method of summing over wavelength represents an upper limit in the sense that individual kinematic components are treated as a single slab. Since the intensity scales with density squared, this estimate will be low by a factor equal to the square root of the number of components. Typically for the total field the emission occupies a mean bandwidth of 4-8Å. If there is one component per thermal broadening width (~2Å) then this underestimates the total by a factor of 1.4-2 or an error of ~0.2 dex.

The next uncertainty is the location of the gas along the line of sight. However, it can be seen from Figure 24 that the radial dependence is much weaker than the dependence on cloud density. This is because the neutral fraction increases with the radius flattening the neutral column density variation. In Figure 27 we show how conversion of intensity to intrinsic density without knowledge of the radial distance leads to a typical error in ~0.3 dex. Also, the thickness of the cloud ($L$) has a modest impact on the results: doubling the thickness produces a mass estimate 0.1 dex higher.

The volume filling factor ($f_V$) has a more significant effect on the mass estimate. All three photon production processes are affected in a similar way by a less then unity $f_V$. For photon pumping, it is easiest to consider the case of an optically thin Lyα line. In this case,

$$I_{PP} \sim nLf_V \left(\frac{n_H}{n}\right) g(\tau_0, b, T).$$

where $L$ is the thickness and $n$ is the local density of the medium in the filled volume. The function $g$ accounts for optical depth and temperature effects. In photoionization equilibrium,

$$\left(\frac{n_H}{n}\right) = \frac{\alpha_{RR}}{\Gamma} n,$$

with so the resulting intensity scales as

$$I_{PP} \sim n^2 Lf_V g(\tau_0, b, T).$$

For the conditions in the gas near the QSO the photoionization rate exceeds the collisional ionization rate ($\Gamma \gg n\gamma$) by >100 so this is a good approximation. This means that for a given observed intensity, the density scales as $n \sim I_{PP}^{1/2} f_V^{-1/2} L^{-1/2}$, and the total mass scales as $M \sim nLf_V \sim \left(I_{PP}^{1/2} f_V^{-1/2} L^{-1/2}\right) Lf_V \sim \left(I_{PP} f_V L\right)^{1/2}$. We find that when non-unity filling factor and different cloud/filament thicknesses are incorporated into the model, including all processes, that the inferred baryon mass is

$$M_{gas} \simeq 3.6 \times 10^{12} f_V^{0.66} \left(\frac{L}{40 kpc}\right)^{0.37} M_\odot.$$

There is no simple way to estimate the volume filling factor without resorting to detailed physical simulations that probe small scales. We might expect $f_V < 1$ in a multiphase CQM, because of the presence of some hot, virialized gas in the massive halo hosting the QSO which can pressure confine cooler clouds responsible for the Lyα emission. Physical processes that might lead to less than unity filling factor in the CQM include turbulent shear between inflowing cold and hot accretion flows, multiphase QSO outflows, or the interaction of inflows, outflows, and orbiting CQM gas. The clouds almost certainly would have to be pressure confined because they are heated dramatically by the QSO and would expand in timescales much less than the QSO lifetime if they are <1 kpc in size. Taking (somewhat arbitrarily) $f_V = 0.3$, the mass estimate is lower by a factor of 2.

The metallicity assumption also impacts the predicted intensity moderately. Consider the behavior at log R[cm]=23.5. If the metallicity is as low as 1%, the inferred density is increased by less than 0.08 dex. If the metallicity is as high as solar, the predicted Lyα flux is decreased and the inferred density increases. This effect is largest at higher density. At log n=-2.0 [-2.5], the inferred density and mass would increase by 0.24 [0.06] dex, or 1.7 [1.14].

The temperature predicted by the CLOUDY model is high. Gas at this temperature would produce very broad absorption (not seen in proximity systems) but because of the low optical depth (typical equivalent widths less than 0.3Å) and width (*b* values of 50-100 km/s) the lines could be difficult to detect. However, if the temperature is in error we could derive an incorrect mass. To check this we fixed the temperature to $T = 10^{4.6} K$. We find that depending on the filling factor the fixed temperature model is lower in mass by a factor of 1.2-1.4.

Finally we can estimate the total uncertainty in the mass estimate based on all these factors. We take a fiducial filling factor of $f_V = 0.3$ and we take the following estimated errors for radial distance, component number, filling factor, thickness, and metallicity as $\sigma_R = 0.3$, $\sigma_n = 0.2$, $\sigma_f = 0.5$, $\sigma_L = 0.3$, $\sigma_T = 0.1$ and $\sigma_Z = 0.1$ (all in dex). Based on the model sensitivities, this yields a total error of $\sigma_M = 0.52 = \left(0.3^2 + 0.2^2 + (0.66*0.5)^2 + (0.36*0.3)^2 + 0.1^2 + 0.1^2\right)$. The baryon mass estimate is then $\log M_{gas} = 12.5 \pm 0.5$ and the total mass is then $\log M_{tot} = 13.4 \pm 0.5$. This estimate is tentative given the simple assumptions of the model. We return now to the two filaments discussed above. Profiles of derived physical parameters are given in Figure 28 and Figure 29. These plots give intensity, intrinsic density *n*, HI column density *N(HI)*, total column density *N(H)*, temperature (from the model), velocity dispersion (from the line profile), and velocity centroid. The figures show the virial radius. It is interesting to note that the velocity dispersion for both filaments is low (~100 km/s) outside the virial radius, but increases inside to 200 km/s (filament 1) and 300-400 km/s (filament 2).

Filament 1 is strongly detected out to 30 arcsec, and possibly detected as far as 45-50 arcsec in the S/N>2.0 adaptively smoothed image corresponding to 2-sigma (~1700LU) with a feature linewidth ~1Å and structure size ~10x10 arcsec$^2$ (see Figure 21). Therefor the filament length is 230-400 kpc. The filament has a baryon mass of $M_{gas} \approx 1.2 \times 10^{11} M_\odot$, baryon mass per unit length of $M_{gas}/l \approx 3 \times 10^8 M_\odot kpc^{-1}$, total mass of $M_{tot} \approx 9.7 \times 10^{11} M_\odot$, and total mass per unit length of $M_{tot}/l \approx 2 \times 10^9 M_\odot kpc^{-1}$ (all assuming unit filling factor, and with errors as estimated above). For filament 2, these parameters are quite similar (length 400 kpc, $M_{gas} \approx 3.6 \times 10^{11} M_\odot$, $M_{gas}/l \approx 8 \times 10^8 M_\odot kpc^{-1}$, $M_{tot} \approx 3 \times 10^{12} M_\odot$, and $M_{tot}/l \approx 7 \times 10^9 M_\odot kpc^{-1}$). We regard the results for filament 2 more tentative because of the uncertainties discussed above.

The spatial morphology of filament 1 and possibly filament 2 may resemble simulations of cold accretion flows (Dekel et al. 2009). In particular we emphasize the low velocity dispersion in the lines, suggesting relatively quiescent gas. Of course the gas has recently been energized by the QSO. The impact of local ionization, heating, and radiation pressure may have not yet had time to disrupt inflowing gas.

There is evidence that filament 1 is an inflow. Filament 1 has a wavelength near 4682Å. The QSO systemic redshift is $z_{sys} = 2.843 \pm 0.001$ or $\lambda_{sys} = 4671.9 \pm 1.2$ Å. The systemic redshift is estimated using the MgII emission line, various Balmer lines, the onset of the Ly alpha forest, and the center of mass of the galaxies in its vicinity, all of which give the same redshift $z_{sys} = 2.843 \pm 0.001$ (Trainor & Steidel 2012). Filament 1 is redshifted with respect to systemic by $650 \pm 80$ km/s, an 8-sigma detection. There is also an absorption feature at this wavelength, but it is dominated by a metal line complex, AlII 1671Å, associated with an intervening Lyα absorber at z=1.802. If a local absorber is present it is masked by the intervening system. The filament is clearly physically associated with the QSO.

Since the emission appears on the red side of the systemic QSO velocity, three models for the filament are possible. The filament could be a tidal stream between the QSO host galaxy and an interacting object. However, the length of filament 1 is as much as an order of magnitude longer than typical tidal streams. The velocity shear between the filament and the QSO is also very large at the base of the filament (Rampazzo et al. 2005).

Alternatively, if the gas is on the far side of QSO it could be outflowing. Extended emission line nebula are common in radio-load QSOs (Heckman et al. 1991a; Heckman et al. 1991b). They however exhibit large velocity dispersions ($\sigma_v$>500 km/s), have scale lengths of ~20 kpc, and do not display the narrow morphology of filament 1. While cooling could in principle lead to a negative velocity dispersion gradient with radius, this behavior the very narrow line widths observed in the distant regions of filament 1 are atypical for any QSO or galactic superwind ouflow. In addition, outflows (or jets) tend to be two-sided. There is no blue-shifted component corresponding to filament 1. Filament 2 is nearly opposite, but has a velocity near systemic. HS 1549+1919 is radio quiet. Radio quiet QSOs are also observed to have extended nebulae (Christensen et al. 2006), with lower velocity dispersions ($\sigma_v$~200 km/s) and large velocity shears with respect to the QSO systemic velocity. Their Lya luminosity is typically ~1% of the QSO luminosity, similar to our object. These do not display narrow filamentary morphology, but previous observations have not had the sensitivity to reach the low flux levels observed by PCWI. They are understood either as outflows or as inflows.

The inflow scenario is the most plausible and least contrived. From Figure 21 and Figure 28 we note that there is a significant rise in the velocity dispersion as the we move along the filament toward the QSO. This rise is similar to that predicted by simulations (P. Hopkins, private communication). Again, it is difficult to understand physically how an outflow would be increasingly collimated and with decreasing velocity dispersion at larger distances, where the dispersion drops to as low as 50 km/s. An outflow with a typical cone angle would exhibit a velocity width larger than observed (comparable to the outflow velocity). The jump in velocity dispersion at R=200 kpc is accompanied by a small redward shift in the velocity centroid of the emission, suggesting an accelerating inward flow. This point is near the clump that is visible in the narrow-band image, region *a* in Figure 1. The infall velocity of ~600 km/s (depending on the angle to the line of site) is consistent with the high dark matter mass derived above for the system. From the derived mass per unit length, and assuming an infall velocity of 600 km/s, we would estimate a mass inflow rate of $\sim 200 M_\odot yr^{-1}$. Detailed geometric, kinematic, and radiative transfer modelling, beyond the scope of this paper, is required to confirm that the observations do not permit alternative interpretations.

### 5.3. Kinematic Properties

The PCWI instrument allows us to study the kinematics of the gas. First we calculate the flux-weighted mean velocity and velocity dispersion of all the extended emission. The distribution in velocity space relative to the QSO systemic velocity is given in Figure 30. The mean is +10 km/s, within errors equal to the systemic velocity. The 1D line-of-sight velocity dispersion is 700 km/s. This is large, corresponding to a 3D velocity dispersion of 1200 km/s, and a halo mass of $1.5 \times 10^{14} M_\odot$, and a factor of ~10 larger than the dark matter mass derived above. We show below that this large velocity dispersion may be an artifact of the superposition of virialized and unvirialized components.

We turn to the next highest moment. The velocity field is complex, but some order can be discerned. Referring back to **Figure 4**, filament 2 is blueshifted toward the west, filament 1 is redshifted toward the south-east, and there is tendancy for the other gas to follow this trend going from blue to red. We can calculate a projected angular momentum, assuming that the gas is in the plane of the sky (Note that if the motion were purely radial the total angular momentum would be zero). The result is $L \approx 10^{16} M_\odot$ km/s kpc, in the S-SW direction (assuming $f_V = 1$). This is consistent with a halo mass of $3 \times 10^{13} M_\odot$ which for a canonical spin parameter λ=0.05 would have an total (baryonic) angular momentum of $1.2 \times 10^{17} (2 \times 10^{16}) M_\odot$ km/s kpc.

We showed earlier how intensity and density are approximately correlated even ignoring distance effects (see Figure 27). Thus it is interesting to consider how the kinematics vary with intensity and therefore density. We might expect that higher density gas is more likely to have collapsed and virialized, while lower density gas may have yet to collapse. We show in Figure 31 the spectra obtained in intensity bins for the full field, along with Gaussian fits to the individual kinematic components. In fact what is apparent is that at low intensities the spectrum is best fit by multiple, separate, narrow components. As the intensity increases components drop out, and a broad component appears centered on the QSO systemic velocity. This trend is summarized in Figure 32 and Figure 33, strongly suggesting that the progression in intensity and density is probing virialization resulting from heirarchical structure formation. The broad central component has a 1D velocity dispersion $\sigma_v \approx 350-400$ km/s, corresponding to a circular velocity of $v_c \approx 600-700$ km/s, and a virial mass of $M_h \approx 2 \times 10^{13} M_\odot$, consistent with the gas mass calculation above.

## 6. Summary

We have used PCWI to detect, map, and characterize the extended Lyα emission in the vicinity of QSO1549+19. We summarize some of the main conclusions here.

1. Extensive emission is detected near the systemic velocity of the QSO, at intensities consistent with expected Lyman α fluorescence in gas of physical densities of $10^{-3}$-$10^{-1}$ cm$^{-3}$, corresponding to overdensities at this redshift of $\delta \sim 10^2 - 10^4$.

2. Extensive simulation tests show that the adaptive smoothing algorithm provides a robust method to detect extended, low surface brightness, filamentary line emission in low to moderate signal-to-noise spectral data cubes.

3. While the emission morphology is complex, clear evidence of filaments are present. In particular a narrow, kinematically cold filament (#1) is detected east of the QSO and redshifted ~10Å with respect to the systemic QSO redshift. There is tentative evidence for a second filamentary or web-like extension west of the QSO connecting the QSO to a previously detected Lyman alpha emitter, which we denote filament 2. These filamentary extensions are detected as far as 50 arcsec, or 400 kpc from the QSO.

4. The faintest regions of emission also display more quiescent kinematics, and both spectral image plots and intensity-binned spectra show a clear trend for faint, distant emission to be kinematically quiet, while emission nearer the QSO and brighter in intensity (and likely density) shows higher velocity dispersions.

5. Evidence is found for an inflowing filament of cold gas (filament #1). The filament is narrow, kinematically cold at the extremity, and progressively disturbed kinematics as it approaches the QSO. As such it is consistent with cold flow accretion models of inflowing gas. Alternate models are less compelling, but a definitive conclusion will require further modelling and observations of additional lines to constrain gas metallicity, radiative transfer, and geometry.

6. A broad central component is detected with a velocity dispersion $\sigma_v \approx 350-400$ km/s, corresponding to a circular velocity of $v_c \approx 600-700$ km/s, and a total virial mass of $10^{13.4\pm0.5} M_\odot$.

7. A simple CLOUDY ionization model is used to deduce physical properties of the gas, notably density and gas mass. Assuming unit filling factor, filament gas masses are $\sim (1-3)\times 10^{11} M_\odot$ while the total gas mass is $\sim 3\times 10^{12} M_\odot$, consistent with the kinematically derived virial mass of $10^{13.4\pm0.5} M_\odot$. We find that to first order density and intensity are monotonically related, independently of distance from the QSO. The absence of extensive regions of double-peaked Lyα emission nominally expected from optically thick Lyα fluorescence is a direct result of the bright radiation field which renders most clouds optically thin in the Lyman continuum with only modest HI column densities.

8. The angular momentum of the emitting gas is shown also to be consistent with the deduced halo mass and a typical spin parameter.

## Acknowledgements

We thank Tom Tombrello and Shri Kulkarni for their support of PCWI. We thank Marty Crabill, Steve Kaye and the staff of the Palomar Observatory for their constant support. Nicole Lingner participated in the observations. We are deeply grateful to Dean Joe Shepard, to the Caltech Counselling Office, and to the family of Daphne Chang for their strength and support. We acknowledge the detailed and helpful comments from the anonymous referee. This work was supported by the National Science Foundation and the California Institute of Technology.

**Tables**

| Region | Area [Arcsec$^2$] | Band | Flux 1 | Flux 2 | Flux 3 | Flux 4 | Flux 5 | Luminosity [erg/s] | S/N |
|---|---|---|---|---|---|---|---|---|---|
| **a** | 100 | 4660-4690Å | 8.2e-16 | 8.2e-18 | 2.7e-19 | 81460 | 2630 | 5.7e+43 | 14.5 |
| **b** | 100 | 4655-4685Å | 8.9e-16 | 8.9e-18 | 2.9e-19 | 87540 | 2820 | 6.1e+43 | 16.0 |
| **c** | 150 | 4660-4680Å | 4.9e-16 | 3.3e-18 | 1.6e-19 | 32370 | 1540 | 3.4e+43 | 6.9 |
| **d** | 150 | 4655-4690Å | 7.5e-16 | 5.0e-18 | 1.4e-19 | 49070 | 1360 | 4.4e+43 | 10.4 |
| **e** | 100 | 4660-4680Å | 3.7e-16 | 3.7e-18 | 1.8e-19 | 36520 | 1740 | 5.1e+43 | 6.9 |
| **tot** | 4450 | 4660-4690Å | 2.0e-13 | 3.6e-18 | 1.2e-19 | 35425 | 1140 | 1.3e+46 | 43.5 |
| **QSO** | | 4660-4690Å | | | | | | 1.0e+48 | |

**Table 1. Derived Fluxes and Detection Significance for Regions Defined in Figure 1.**
Flux 1 = erg cm$^{-2}$ s$^{-1}$. Flux 2 = erg cm$^{-2}$ s$^{-1}$ arcsec$^{-2}$. Flux 3 = erg cm$^{-2}$ s$^{-1}$ arcsec$^{-2}$ Å$^{-1}$. Flux 4 = LU. Flux 5 = LU/Å. Errors in each unit can be derived by dividing flux by S/N.

**Figures**

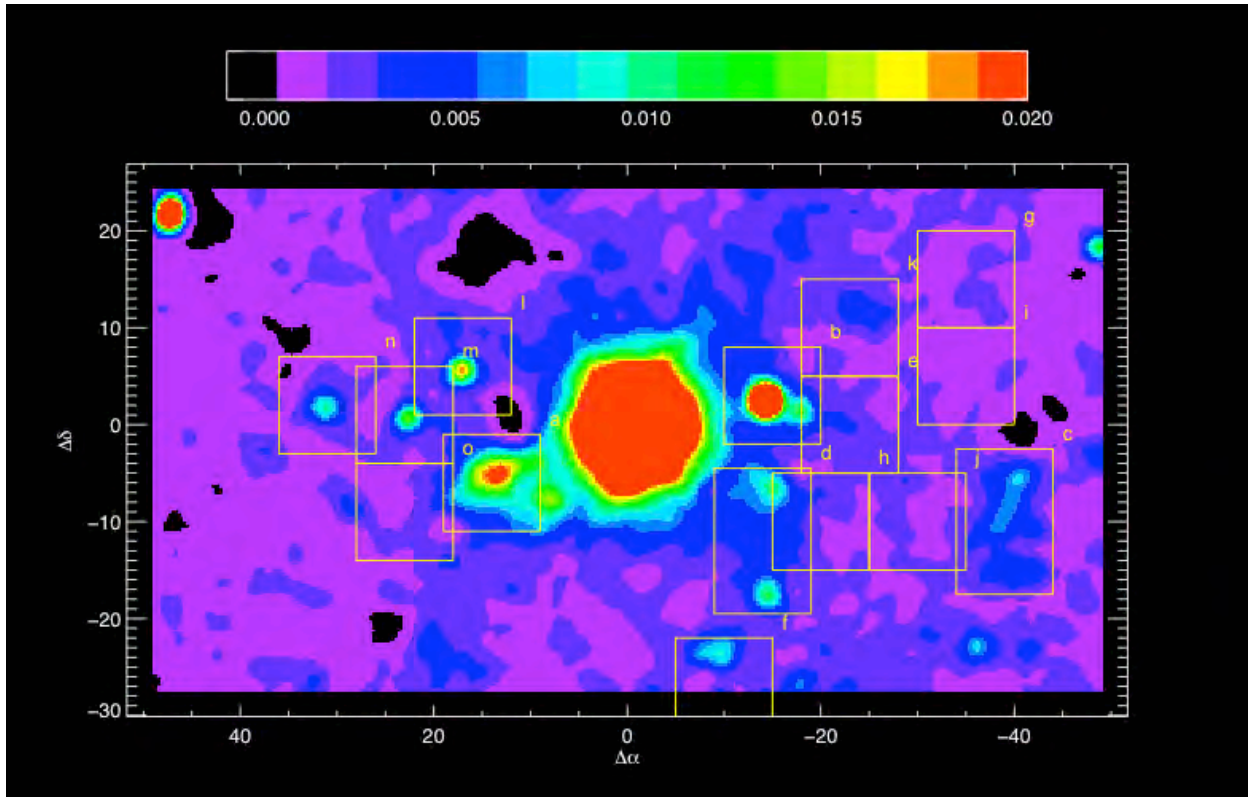

**Figure 1. Images of QSO HS1549+19.**

**a. Narrow band image of QSO HS1549+19.** The image is obtained on LRIS at Keck 1, using a narrow band filter 80Å centered on 4670Å. The image is a co-added sum totaling 3 hours of exposure in the narrow-band, and the V-band image in Figure 16 is used to continuum-subtract the image. The image has been smoothed with a Gaussian kernel of radius 1.5 arcsec. Full scale (0.02) corresponds to a flux of 1.6e-17 erg cm$^{-2}$ s$^{-1}$ arcsec$^{-2}$. The 1 sigma noise is ~2.5e-19 erg cm$^{-2}$ s$^{-1}$ arcsec$^{-2}$ in a 1 arcsec$^2$ box. The continuum subtraction error is estimated to be ~2.5% full scale or 4e-19 erg cm$^{-2}$ s$^{-1}$ arcsec$^{-2}$. We show a number of regions as labeled yellow boxes. The first four (a-d) are distinct "blob" features in the narrow-band image. Each has a LRIS spectroscopic observation and an associated mean redshift.

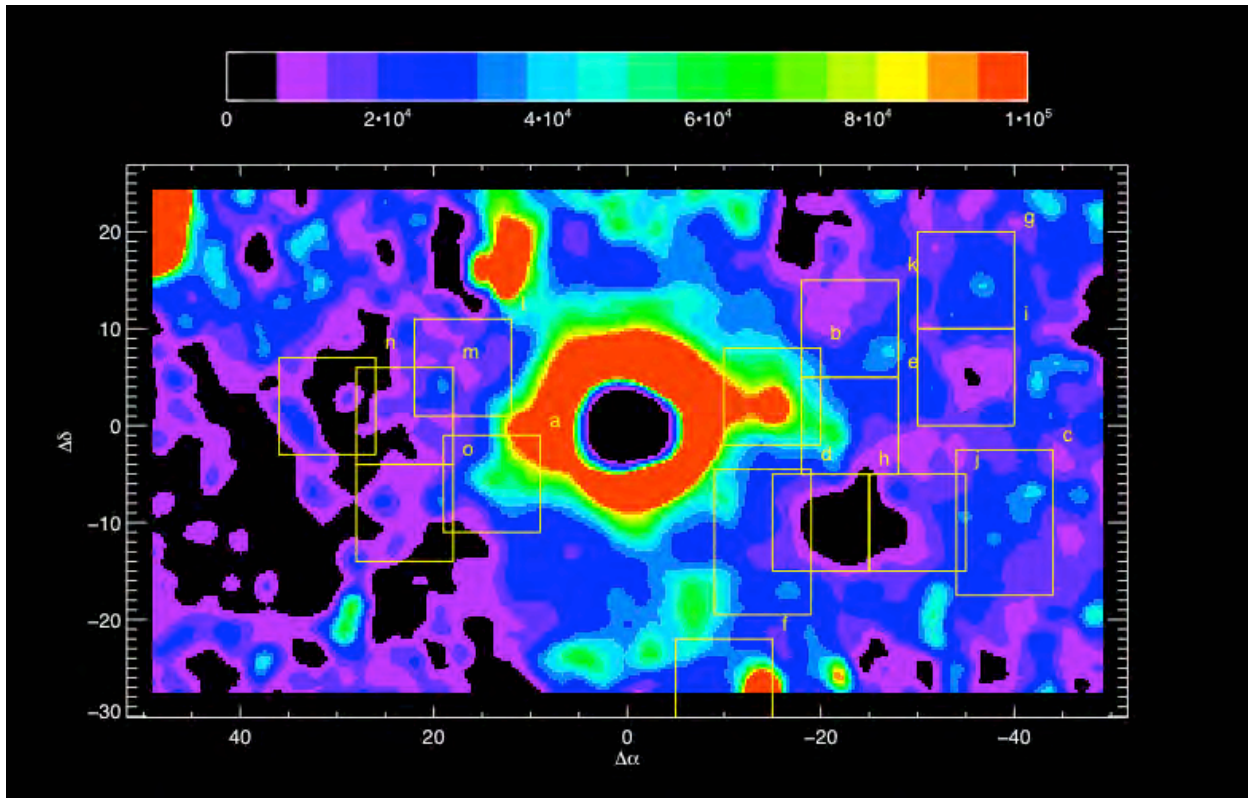

**b. Corresponding PCWI image**, centered on 4670Å, 40Å wide total slice using 3DASL images (see text). The color bar gives a linear intensity scale with 100,000 LU full range. 100,000 LU = 100,000 ph cm$^{-2}$ s$^{-1}$ sr$^{-1}$ = $10^{-17}$ erg cm$^{-2}$ s$^{-1}$ arcsec$^{-2}$. Smoothing is performed using 3DASL with a S/N threshold of 2.5 after masking the QSO within a 5 arcsec radius. Blobs a, b, and c are detected in PCWI image, albeit at lower spatial resolution due to the 2.4 arcsec slice sampling resolution. Blobs d are not clearly detected. Note there are some discrepant regions when comparing panels a. and b. One to one correspondence is not expected because of the lower angular resolution of PCWI, the higher noise in a wide PCWI slice (due to lower throughput, smaller telescope aperture, and higher sky background), and the continuum subtraction error in the narrow-band image. We show in Figure 2 that the two data sets are reasonably consistent.

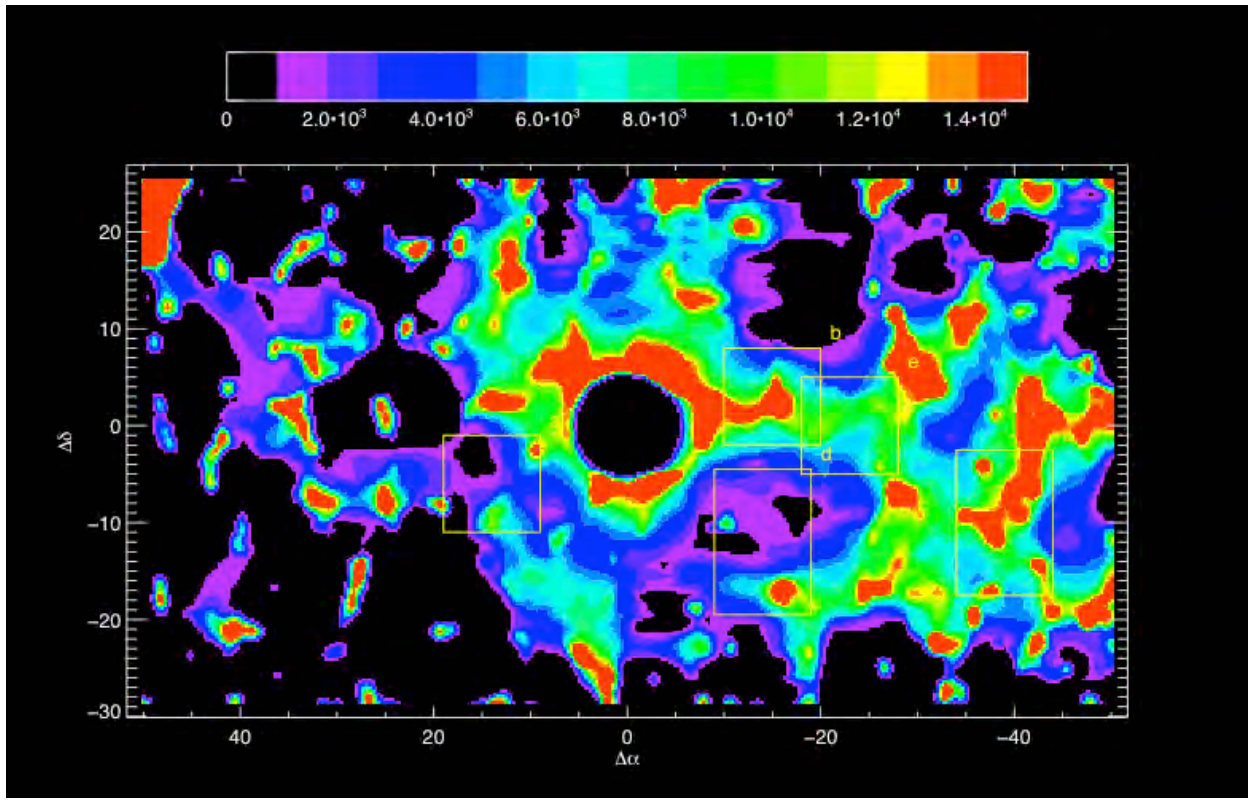

**c. Smoothed spectral images of QSO HS1549+19,** 4Å wide slice, centered on 4672Å. The scale in arcsec is centered on the QSO. The color bar gives a linear intensity scale with 15000 LU full range. 10000 LU = 10000 ph cm$^{-2}$ s$^{-1}$ sr$^{-1}$ = $10^{-18}$ erg cm$^{-2}$ s$^{-1}$ arcsec$^{-2}$. Smoothing is performed using 3DASL with a S/N threshold of 2.5 after masking the QSO within a 5 arcsec radius. Note that the image looks quite different from that in panel b, a result of the much smaller wavelength range near the systemic velocity. In these images we can identify one possible 'filamentary' extension, connecting regions b, e, and c. We tentatively denote this feature "Filament 2".

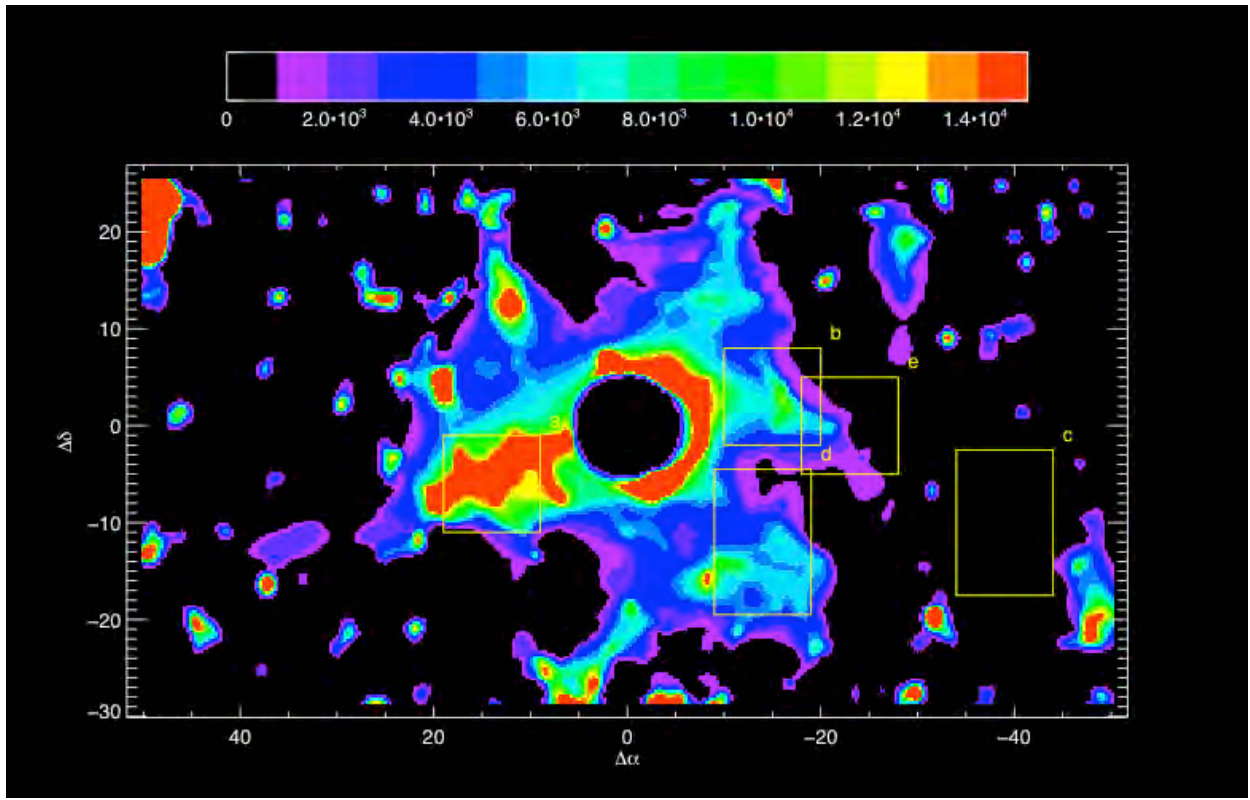

**d. Smoothed spectral images of QSO HS1549+19,** 4Å wide slice, centered on 4681Å. The scale in arcsec is centered on the QSO. The color bar gives a linear intensity scale with 15000 LU full range. 10000 LU = 10000 ph cm$^{-2}$ s$^{-1}$ sr$^{-1}$ = 10$^{-18}$ erg cm$^{-2}$ s$^{-1}$ arcsec$^{-2}$. Smoothing is performed using 3DASL with a S/N threshold of 2.5 after masking the QSO within a 5 arcsec radius. Note that the image looks quite different from that in panel b. What appeared to be a distinct blob in panel a and b manifests as a linear extension eastward, which we denote "Filament 1".

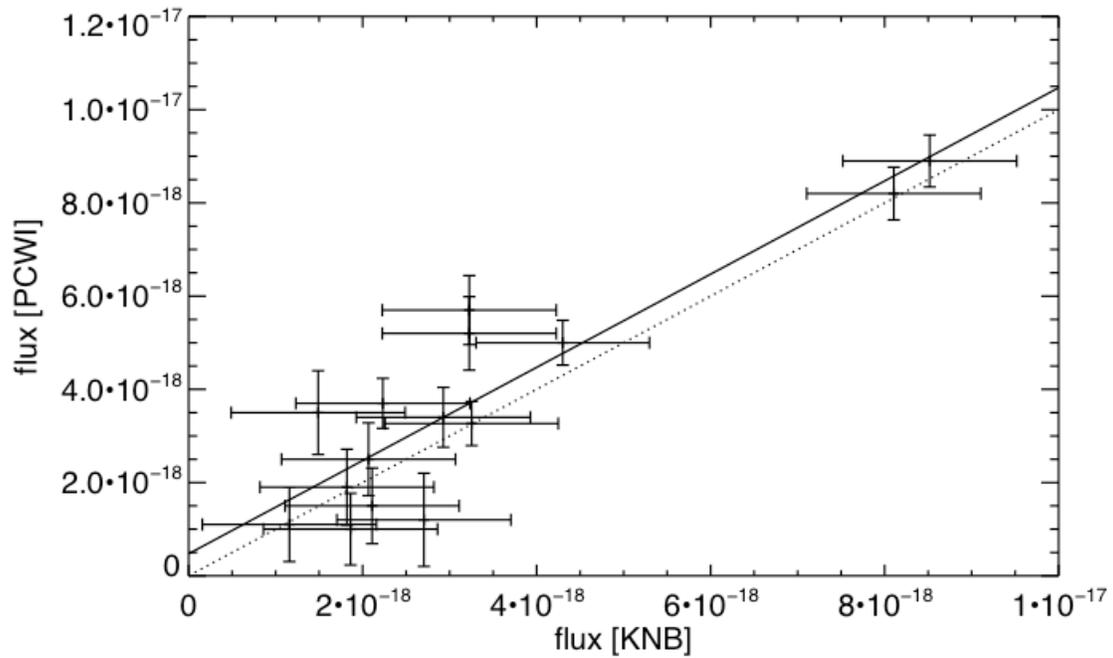

Figure 2. Comparison of flux in regions shown in Figure 1 from the Keck narrow-band image and PCWI shows good correspondence. Both fluxes are in units of erg cm$^{-2}$ s$^{-1}$ arcsec$^{-2}$ averaged in the regions. Error bars for the narrow-band image include estimated continuum subtraction error of ~8 x 10$^{-19}$ erg cm$^{-2}$ s$^{-1}$ arcsec$^{-2}$. Dotted line shows 1 to 1, solid line shows linear fit. Flux offset is 4.7 x 10$^{-19}$ erg cm$^{-2}$ s$^{-1}$ arcsec$^{-2}$. Fit chi-square is 12.8 for 13 degrees of freedom.

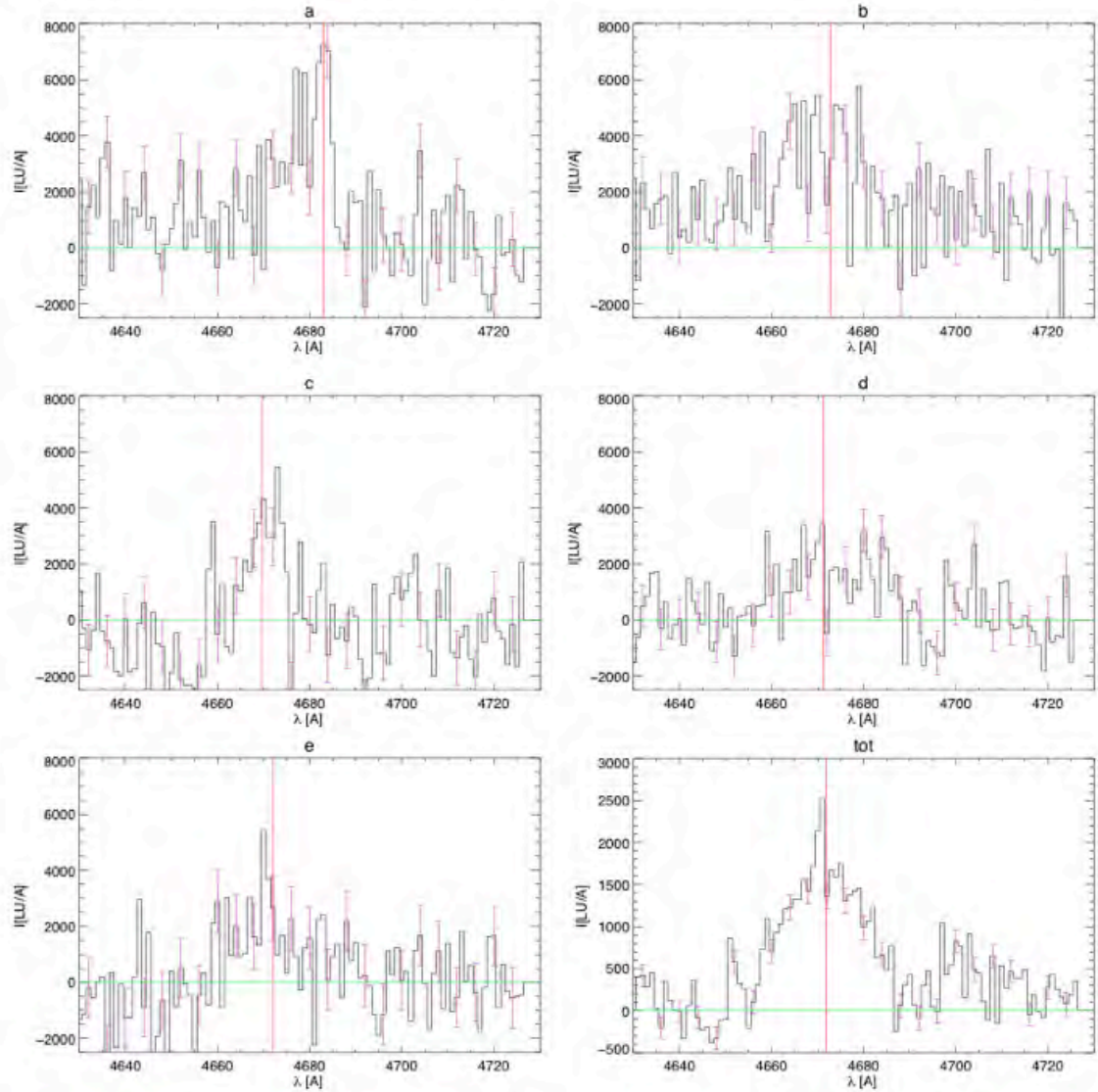

**Figure 3. Spectra from unsmoothed data cube in regions shown in Figure 1.** Units are LU/Å. Error bars are calculated from the variance cube. Red lines show measured redshift based on LRIS spectra. For region e there is no LRIS spectrum, and we show Lyα at the systemic velocity. Regions a and e have quite narrow line widths, region c intermediate, and regions b and d broad. The final panel, 'tot', shows the total spectrum from the QSO region with the QSO masked. Detected fluxes and significance are presented in Table 1.

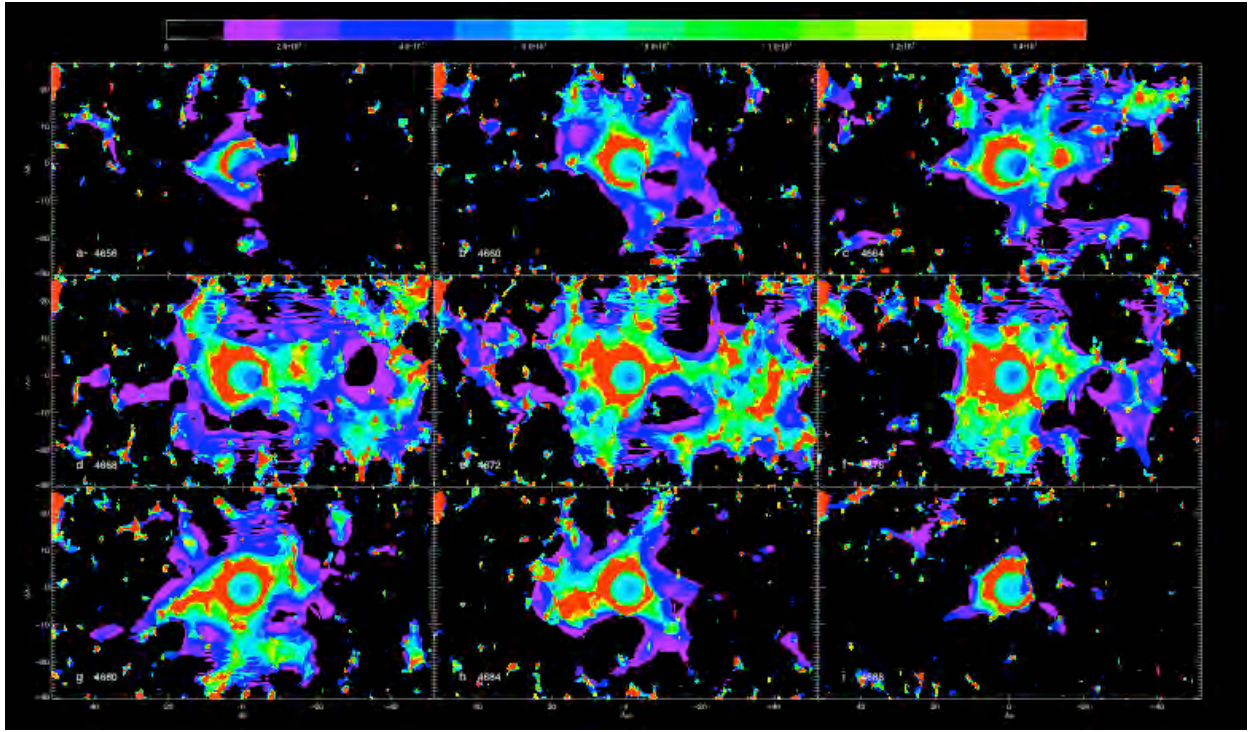

**Figure 4. Smoothed spectral images of QSO HS1549+19 for several wavelengths.** Each panel consists of a 4Å wide slice, centered on a) 4656Å, b) 4660Å, c) 4664Å, d) 4668Å, e) 4672Å, f) 4676Å, g) 4680Å, h) 4684Å, and i) 4688Å. The scale in arcsec is centered on the QSO. The color bar gives a linear intensity scale with 15000 LU full range. 10000 LU = 10000 ph cm$^{-2}$ s$^{-1}$ sr$^{-1}$ = 10$^{-18}$ erg cm$^{-2}$ s$^{-1}$ arcsec$^{-2}$. Smoothing is performed using 3DASL with a S/N threshold of 2.5 after masking the QSO within a 5 arcsec radius. In these images we can identify at least two 'filamentary' extensions. One extends to the south-east in the 4680-4684Å slices, and we denote this "Filament 1". The other connects the QSO with the clump at 4672Å, Δα=-40, Δδ=-7 arcsec that is evident in Figure 1, which we denote "Filament 2.

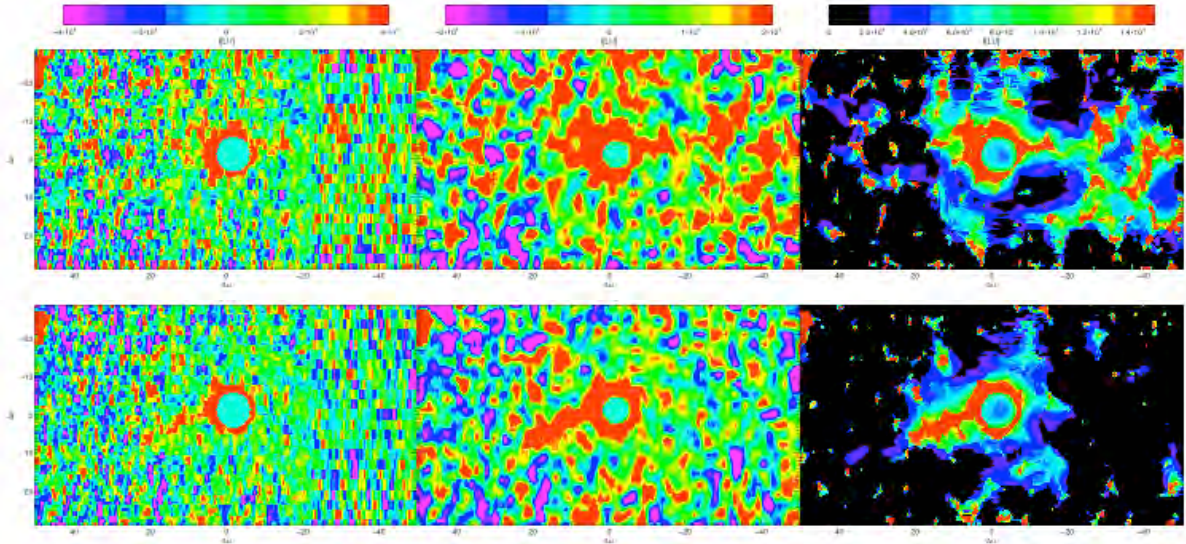

**Figure 5. Comparison of raw and smoothed slices with 3DASL images.** Top row: 4671Å, bottom row: 4682Å. Left: raw difference cube, 4Å slice. Middle: raw difference cube, 4Å slice, smoothed by 9 pixels. Right: 3DASL with S/N threshold 2.5. Note that image scales, given by color bars, are different for each image type.

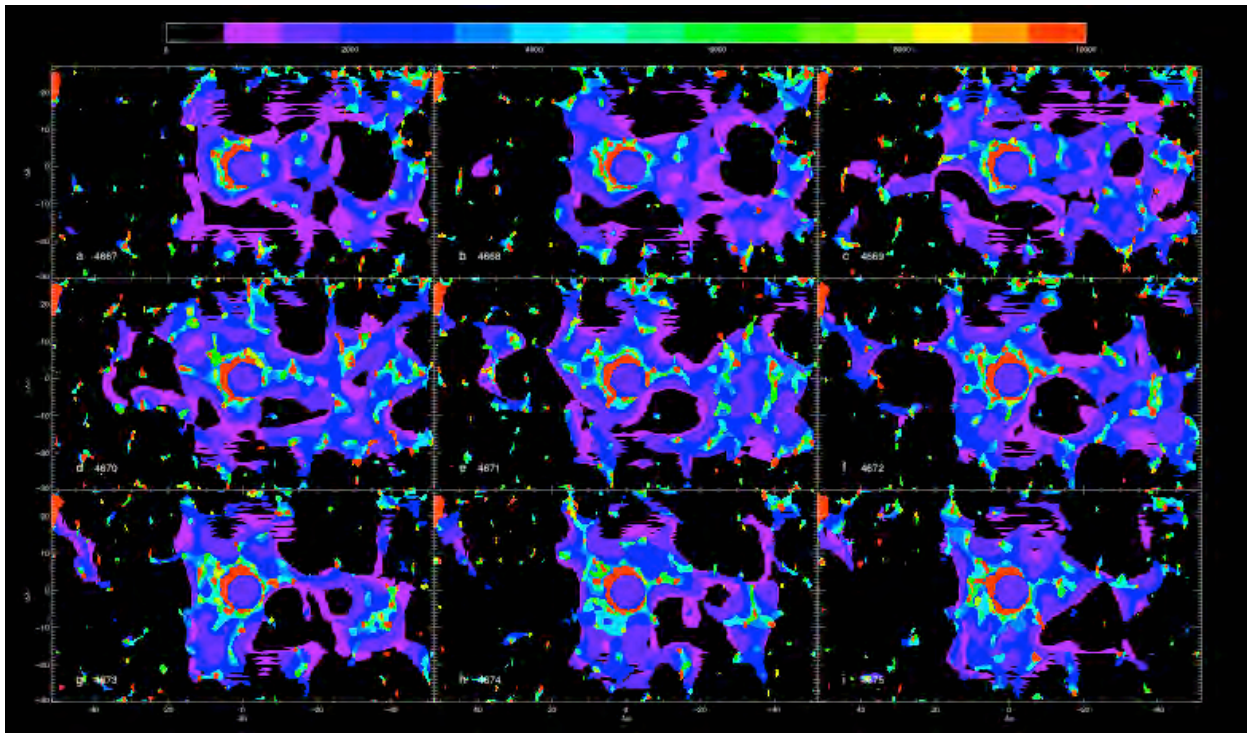

**Figure 6. Smoothed 1Å slices using 3DASL covering band around QSO systemic velocity, 4672Å.** S/N threshold is 2.5. Filament 2 is detected in at least 4 slices.

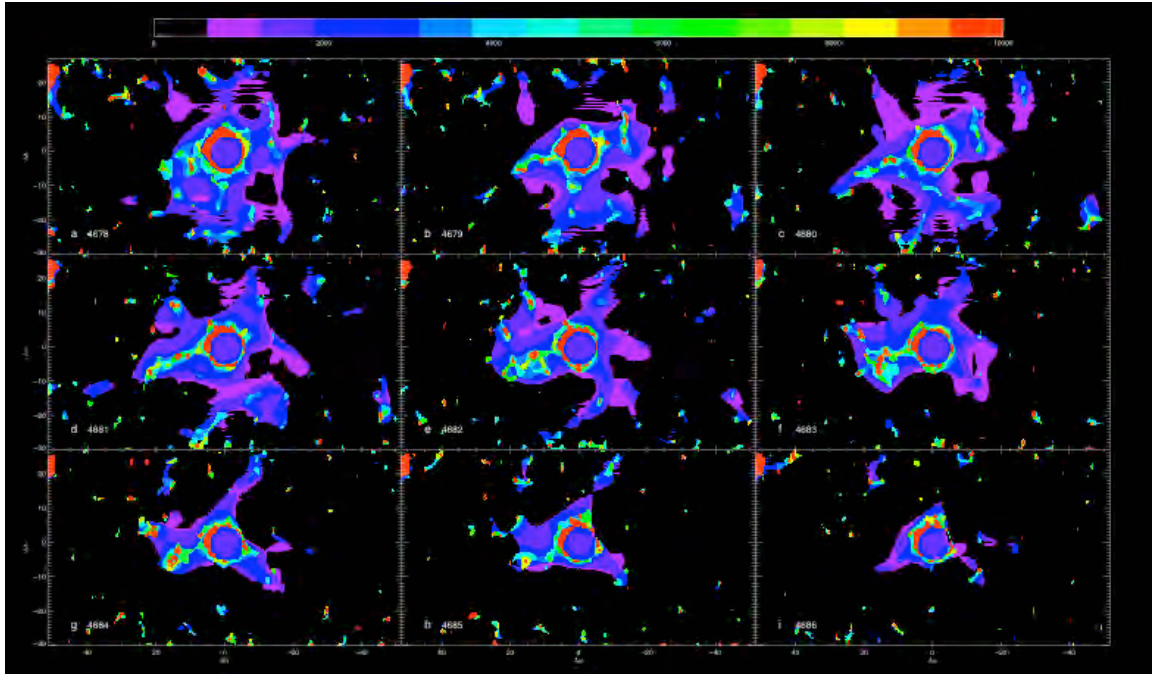

**Figure 7. Smoothed 1Å slices using 3DASL covering band around 4682Å.** S/N threshold is 2.5. Filament 1 is detected in at least 7 slices.

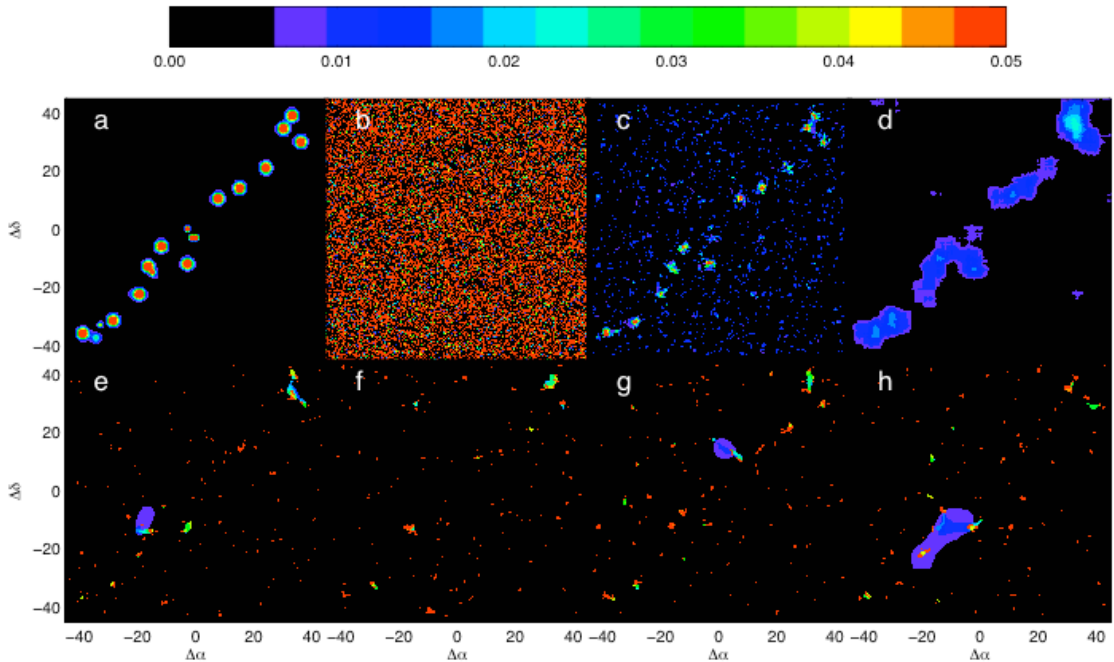

**Figure 8.** Simulated 2D image field with sources only, illustrating several realizations of 2DASL compared to simple smoothing. a. simulated field, no noise. b. simulated field, with noise (one realization, $\sigma=0.2$). c. average of 10 2DASL realizations (example realizations shown in e-h). d. average of 10 simple 30 pixel smooth realizations. e, f, g, h. 4 examples of realizations of 2DASL. Comparison of panels c and d show that simple smoothing generates spurious filamentary features from compact sources distributed in a filamentary pattern, while 2DASL does avoids this problem.

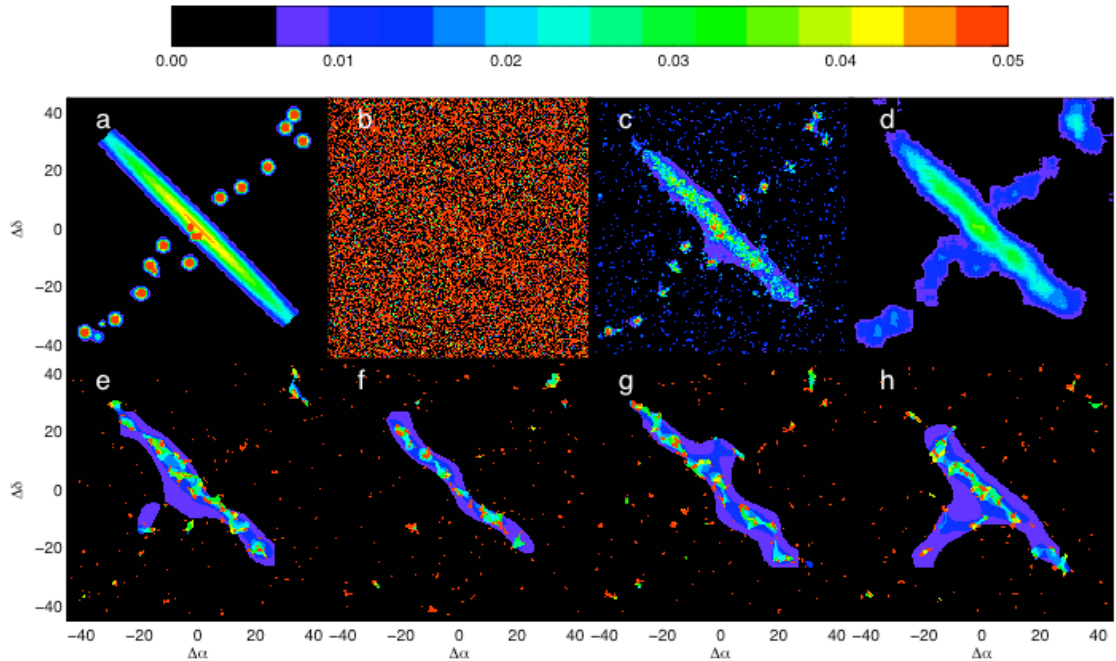

**Figure 9.** Simulated 2D image field with sources and a filament, illustrating several realizations of 2DASL compared to simple smoothing, using a S/N threshold of 2.5. The filament is a 2D Gaussian with FWHM 20 pixels by 300 pixels, with a mean surface brightness (within FWHM) of 0.029.  a. simulated field, no noise. b. simulated field, with noise (one realization, $\sigma$=0.2). c. average of 10 2DASL realizations (e-h?). d. average of 10 simple 30 pixel smooth realizations. e, f, g, h. 4 examples of realizations of 2DASL. As in **Figure 8**, comparison of panels c and d show that simple smoothing generates spurious filamentary features from compact sources distributed in a filamentary pattern, while 2DASL does avoids this problem. Comparison of panels c and d show that simple smoothing generates spurious filamentary features from compact sources distributed in a filamentary pattern, while 2DASL does avoids this problem.

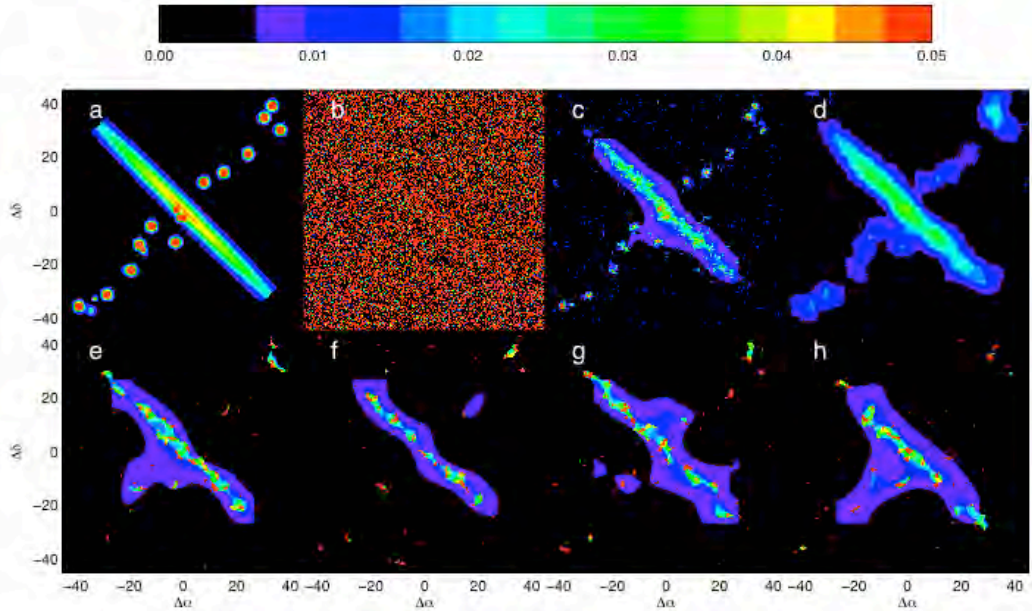

**Figure 10.** Simulated 2D image field with sources and a filament, illustrating several realizations of 2DASL compared to simple smoothing, using a S/N threshold of 3.0. The filament is a 2D Gaussian with FWHM 20 pixels by 300 pixels, with a mean surface brightness (within FWHM) of 0.029. a. simulated field, no noise. b. simulated field, with noise (one realization, σ=0.2). c. average of 10 2DASL realizations (e-h). d. average of 10 simple 30 pixel smooth realizations. e, f, g, h. 4 examples of realizations of 2DASL. See comments in **Figure 8** and **Figure 9**.

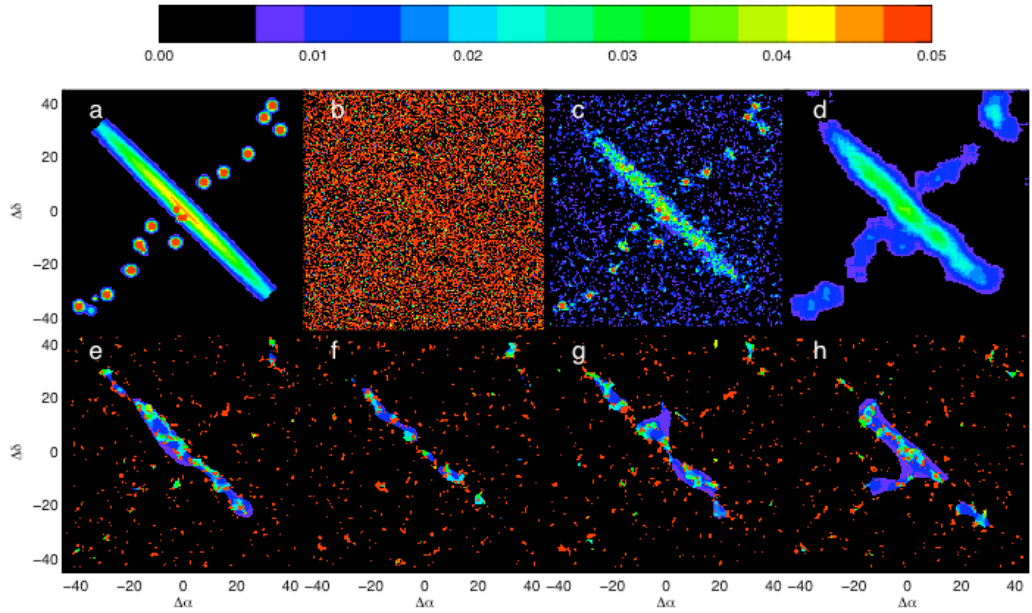

**Figure 11.** Simulated 2D image field with sources and a filament, illustrating several realizations of 2DASL compared to simple smoothing, using a S/N threshold of 2.0. The filament is a 2D Gaussian with FWHM 20 pixels by 300 pixels, with a mean surface brightness (within FWHM) of 0.029. a. simulated field, no noise. b. simulated field, with noise (one realization, σ=0.2). c. average of 10 2DASL realizations (e-h). d. average of 10 simple 30 pixel smooth realizations. e, f, g, h. 4 examples of realizations of 2DASL. See comments in **Figure 8** and **Figure 9**.

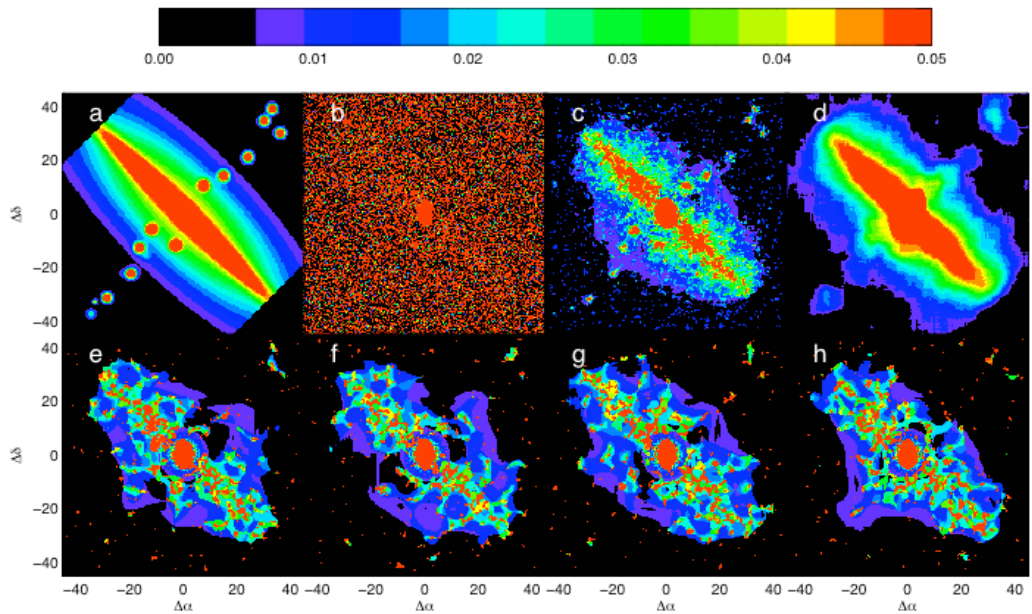

**Figure 12.** Simulated 2D image field with sources (including a very bright source at the center, approximately 10,000 times the surface brightness of the filament, and not masked), and a filament with a narrow and broad component, illustrating several realizations of 2DASL compared to simple smoothing, using a S/N threshold of 2.5. a. simulated field, no noise. b. simulated field, with noise (one realization, σ=0.2). c. average of 10 2DASL realizations (e-h). d. average of 10 simple 30 pixel smooth realizations. e, f, g, h. 4 examples of realizations of 2DASL. See comments in **Figure 8** and **Figure 9**.

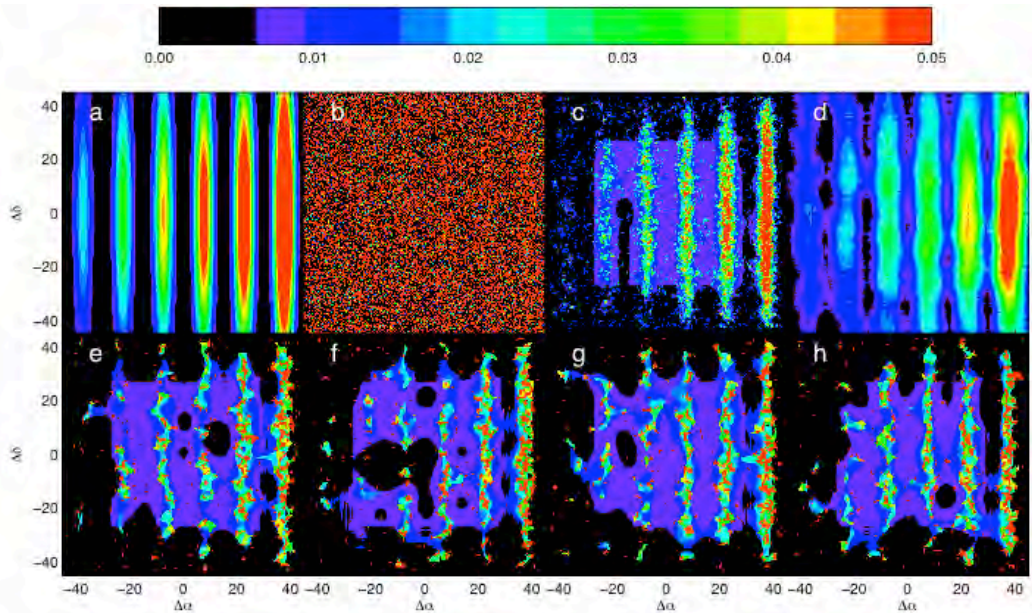

**Figure** 13**. Simulated 2D image field, showing a series of filaments with increasing intensity.** Each filament is a 2D Gaussian with FWHM 20 pixels by 300 pixels. Mean surface brightness (within FWHM) from left to right is 0.014, 0.021, 0.029 (same as Figures 7-9) **Figure 9**, 0.036, 0.043, 0.057. a. simulated field, no noise. b. simulated field, with noise (one realization, σ=0.2). c. average of 10 2DASL realizations (d-g). d. average of 10 simple 30 pixel smooth realizations. e, f, g, h. 4 examples of realizations of 2DASL. 2DASL realizations use S/N threshold of 2.5. Filament 3 (average brightness 0.029) is considered barely detected, as expected since the signal-to-noise ratio is approximately 0.029/(0.2/FWHM)=0.029/(0.2/20)=2.9.

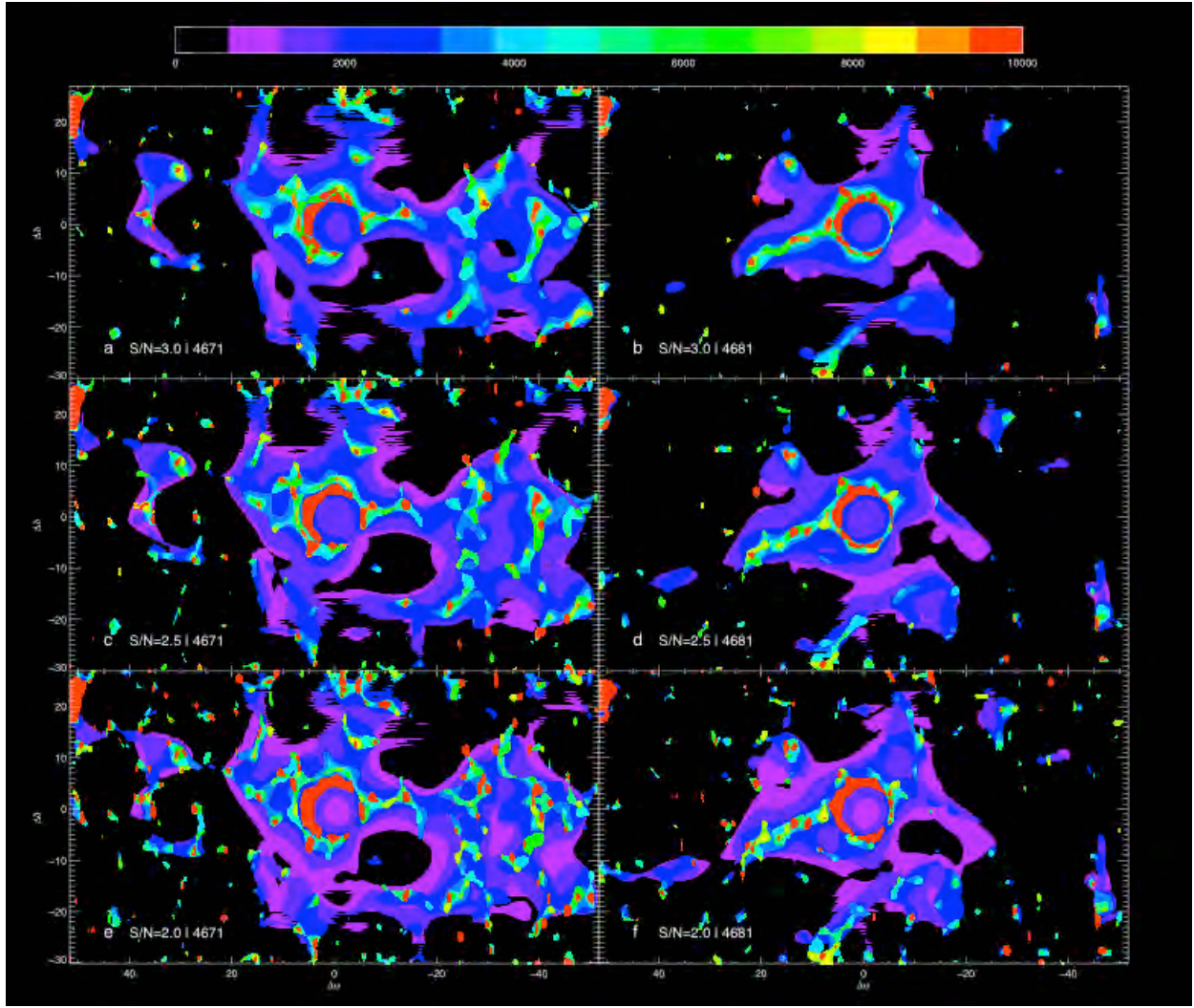

**Figure 14. Smoothed spectral images (1Å slices) for showing the effect of varying the S/N threshold of the algorithm.** a. S/N 3.0, 4671Å. b. S/N 3.0, 4681Å. c. S/N 2.5, 4671Å. d. S/N 2.5, 4681Å. e. S/N 2.0, 4671Å. f. S/N 2.0, 4681Å.

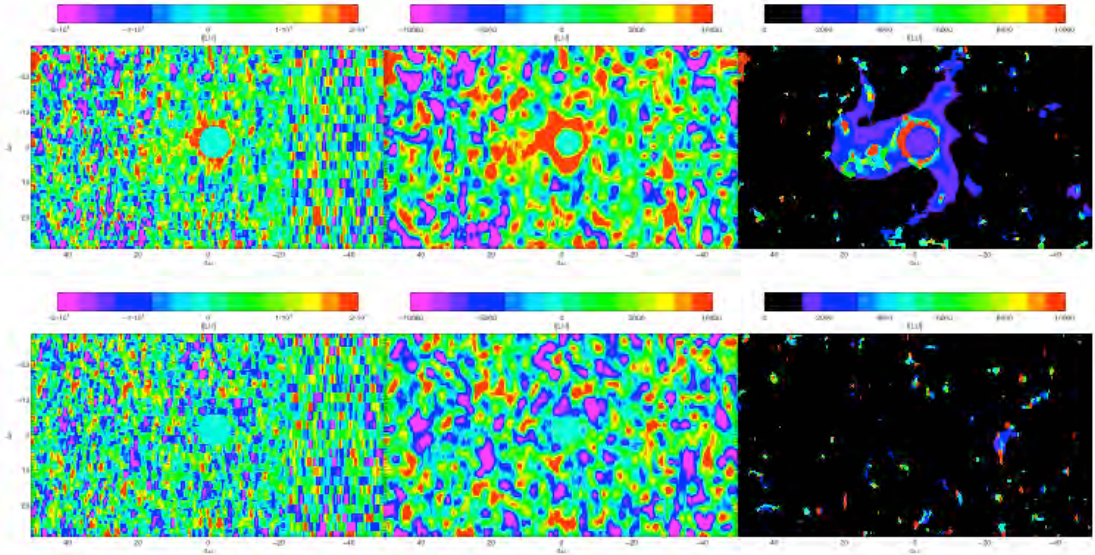

**Figure 15. Raw, smoothed, and 3DASL images of data (top row) in 1Å slice, and pure noise (bottom row) obtained using actual pipeline and image positions in final mosaic, also a 1Å slice.** Left: raw image mosaic. Middle: raw mosaic smoothed by 9 x 9 boxcar. Right: 3DASL image, S/N threshold 3.0, 1Å wide. Note that because of the 3D smoothing, some features that appear in a single 1Å slice will not be detected if they are not correlated from slice to slice. For example, in the data (top row), emission appears at Δα=-30, Δδ=-8, but is absent in the 3DASL image on the right.

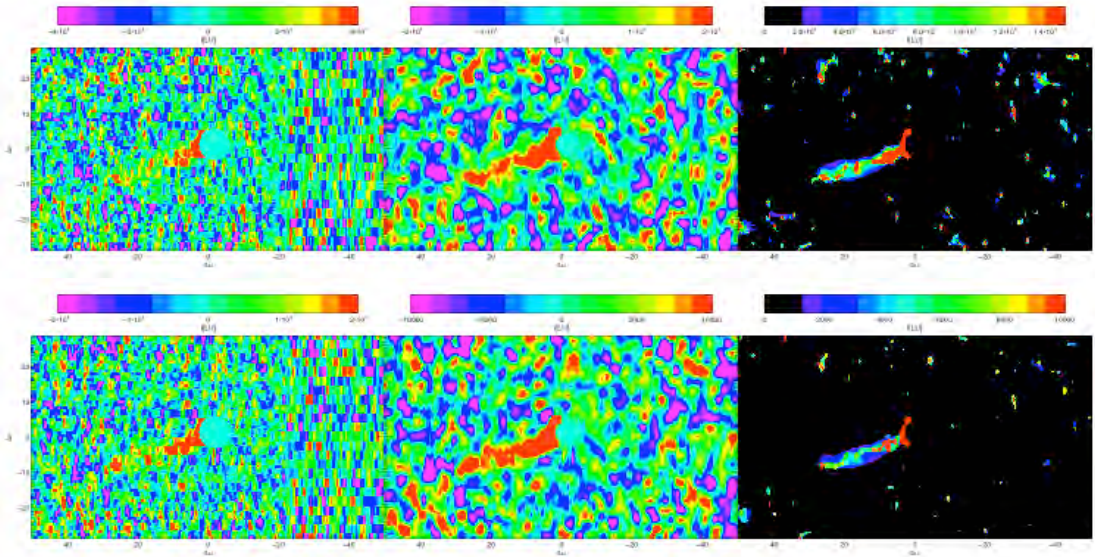

**Figure 16. Raw, smoothed, and 3DASL images of simulated data cube with single line filament similar to observed filament 1, FWHM 3Å, fixed center (4682Å).** Upper row: 4Å slices. Lower row: 1Å slices. Left: raw image mosaic. Middle: raw mosaic smoothed by 9 x 9 boxcar. Right: 3DASL image, S/N threshold 3.0.

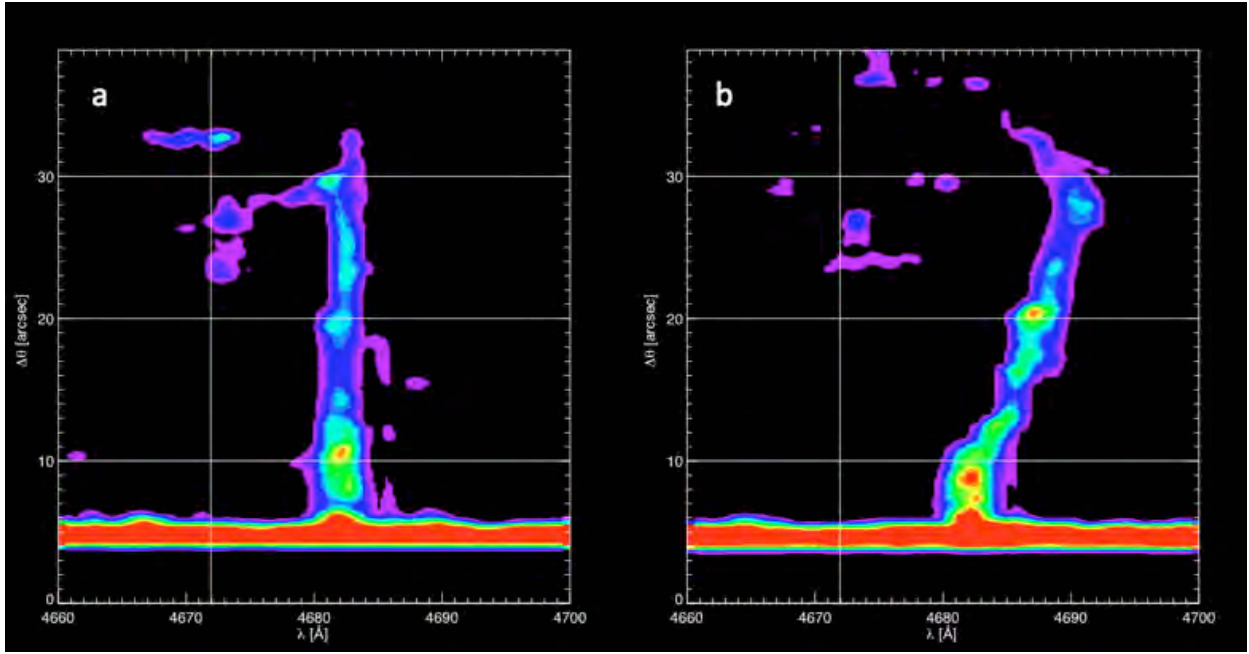

**Figure 17. Simulated data cube with single line filament, smoothed with 3DASL, cut in spectral-image direction along filament, S/N threshold 3.0.** a. No velocity change along filament. b. velocity change ~600 km/s over filament.

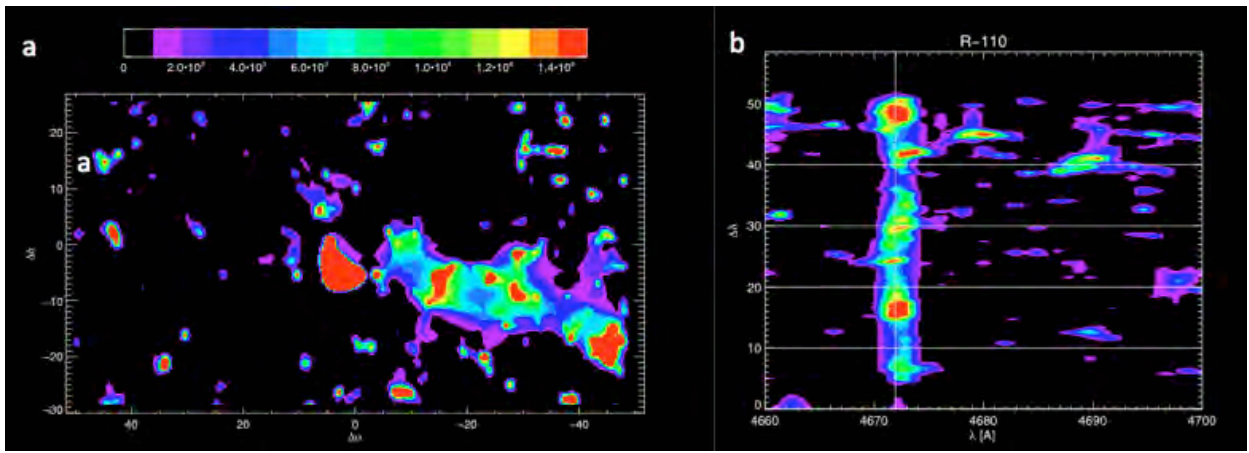

**Figure 18. Simulated filament in western direction of comparable intensity to filament 2.** This simulation demonstrates that the detectability of filamentary emission is not affected by the lack of north-south dithering present in the western portion of the data cube mosaic. a. Image slice centered on 4672Å, 4Å wide from 3DASL cube at S/N>3.0. b. Spectral image plot at azimuth -110°. Filament has total intensity 18,750LU, width 12 arcsec, velocity dispersion 60 km/s.

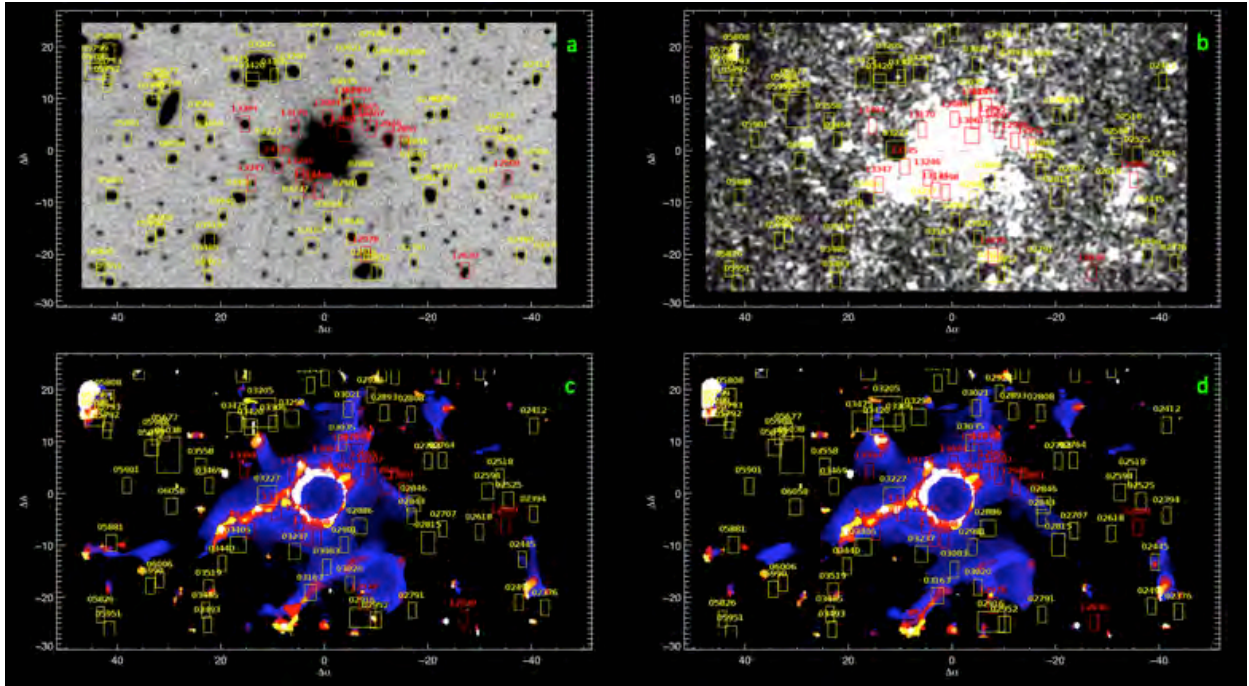

**Figure 19. Illustration of location and impact of compact sources.** a. V-band continuum image showing all compact sources with V<25.0. Yellow boxes show continuum-only sources, and red boxes show Lya emitters. Boxes denote region replaced with local sky on a wavelength by wavelength basis. Sky is obtained using an annular zone. b. Narrow-band image with source boxes. c. 3DASL smoothed 1Å slice at 4682Å after source removal. d. 3DASL smoothed 1Å slice at 4682Å with no source removal. There is no significant difference between the source-removed and non-sourced-removed images in this and at all wavelengths.

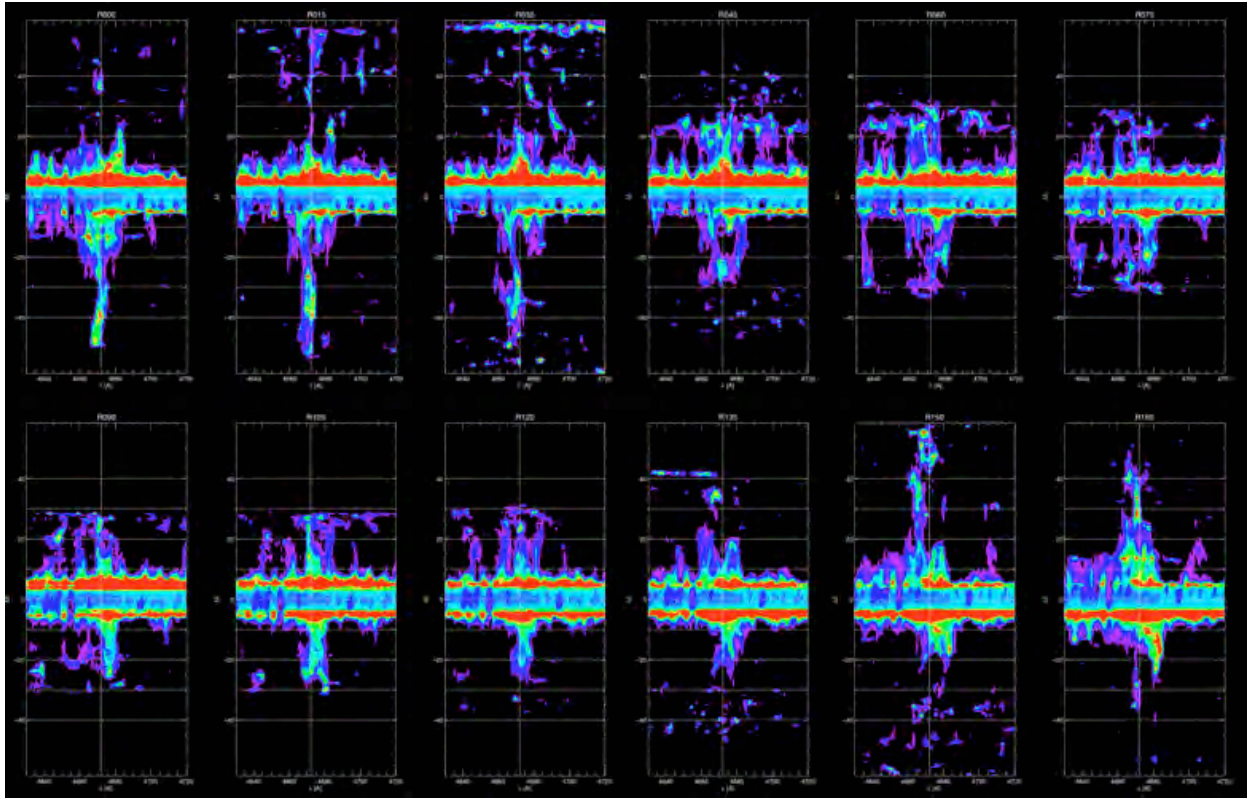

**Figure 20. A set of spectral-image plots at different azimuthal positions centered on the QSO.** Each plot is obtained in a pseudo-slit 5 arcsec wide. R000 is 0 degrees from the horizontal axis towards East. Angle is clockwise from this, so that for example 90 degrees corresponds to a North-South pseudo-slit. Vertical lines corresponds to Lyman α at QSO systemic redshift.

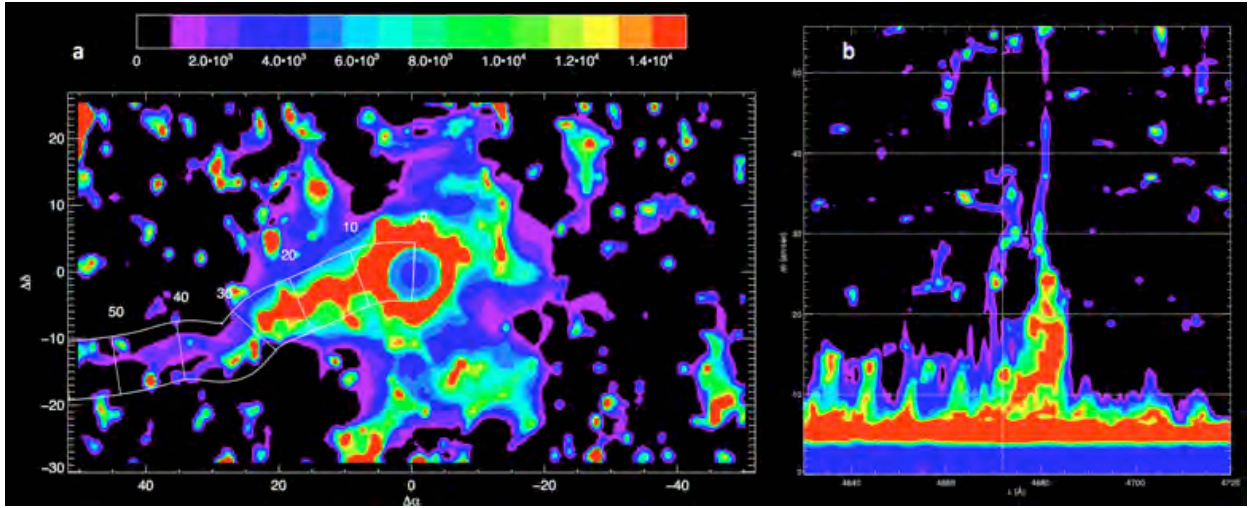

**Figure 21. Smoothed image slices with delineated filament regions for filament 1 and spectral-image plot.** a: 3DASL smoothed image slice, 4Å wide, S/N threshold 2.0 to bring out faintest extension of. b. Spectral image plot, summed over pseudo slit shown in panel a., full scale 5000 LU/Å, ordinate is distance along region in arcsec from QSO. There is a suggestion of a double-peaked line profile in the range Δθ~20-40 arcsec.

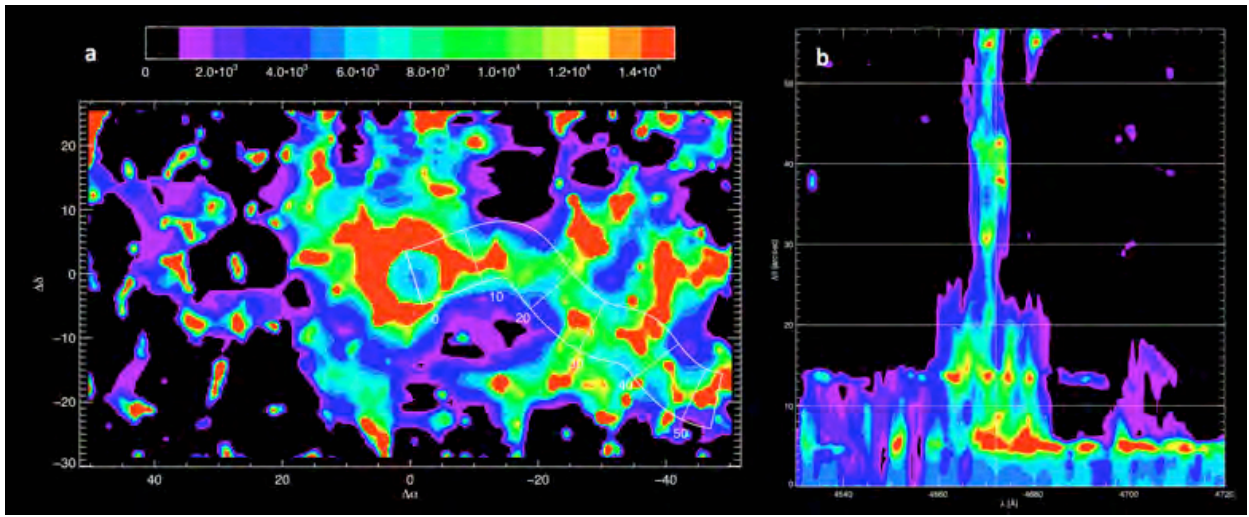

**Figure 22. Smoothed image slices with delineated filament regions for filament 2 and spectral-image plot.** The path of the pseudo-slit chosen to delineate filament 2 is not unique, and other choices produce similar results. a: 3DASL smoothed image slice, 4Å wide, S/N threshold 2.5. b. Spectral image plot, summed over pseudo slit shown in panel a., full scale 5000 LU/Å, ordinate is distance along region in arcsec from QSO.

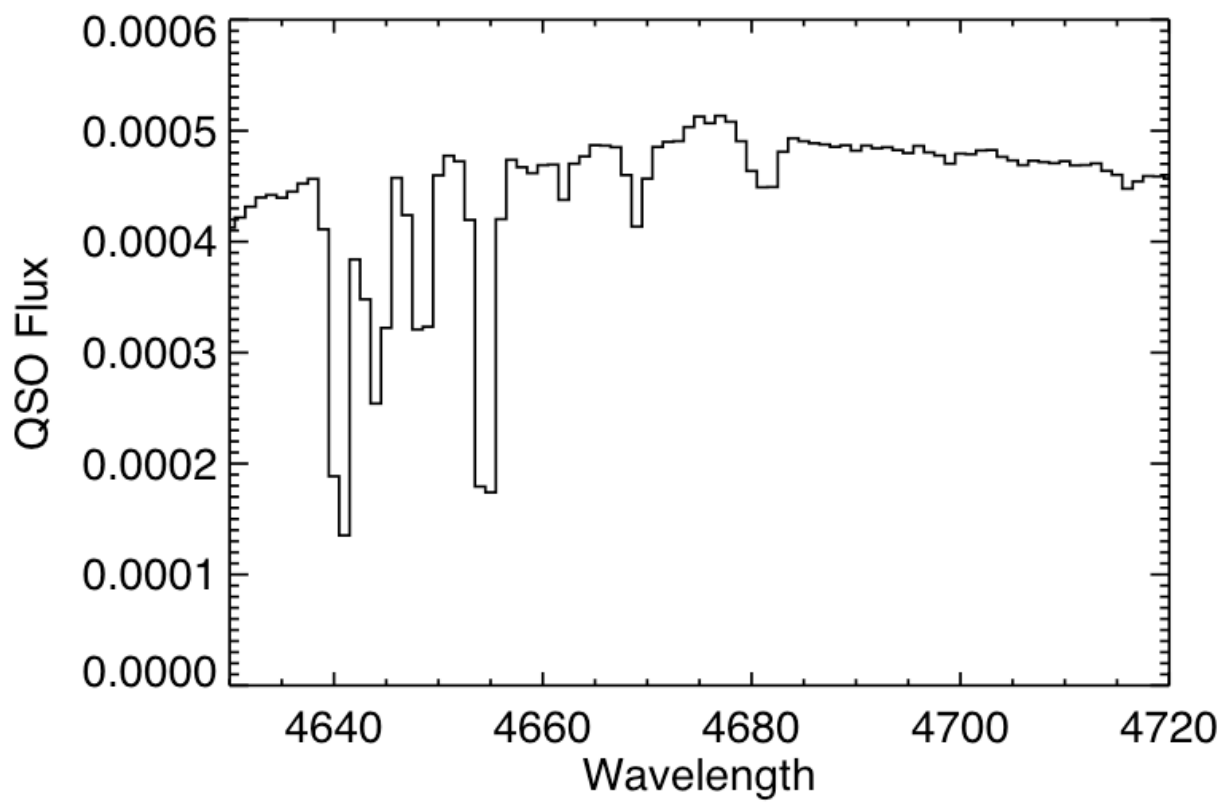

**Figure 23. Spectrum of QSO HS1549+19 in PCWI spectral range.** Absorption lines are present at 4640.5, 4643.9, 4648.3, 4654.4, 4661.9, 4668.9, and 4981.3Å.

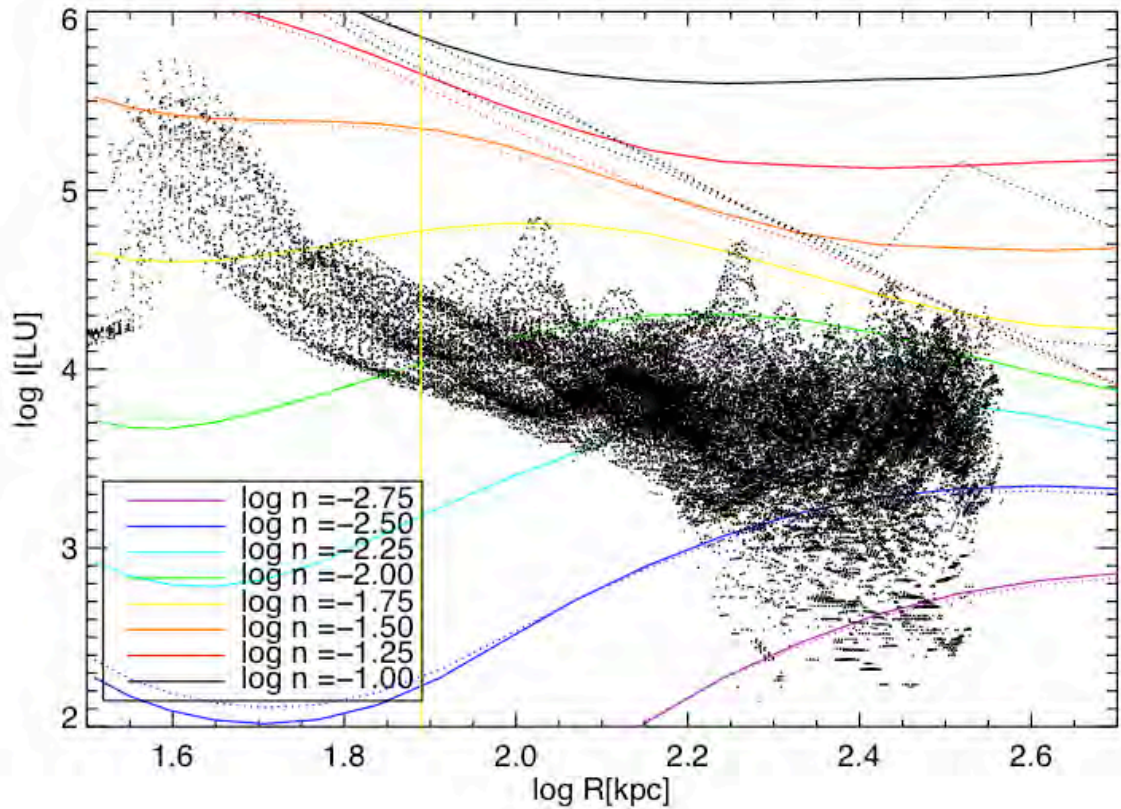

**Figure 24. Inferred cloud density vs. radius and intensity from CLOUDY model.** Dots give measured intensity vs. radius. Colored lines give predicted intensity vs. radius for given densities shown in legend. Solid line gives total intensity, and dotted line gives line-pumped intensity. Only for the highest densities at large radii are significant Lyman α fluorescence and corresponding double-peaked emission from an optically thick source expected. All model results are for volume filling factor unity ($f_v = 1$) and cloud thickness 40 kpc. Yellow vertical line indicates radii within which QSO masking could impact results.

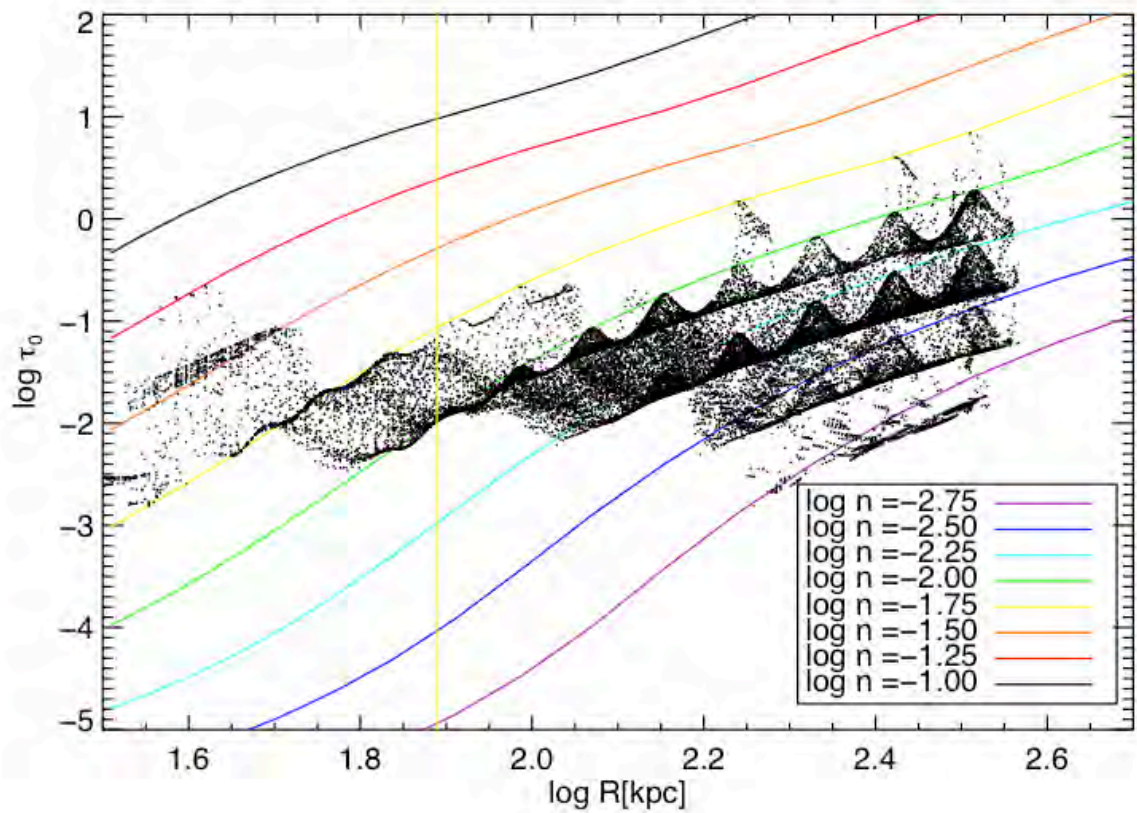

**Figure 25. Inferred cloud line center optical depth vs. cloud density.** Points display intensities converted into line center optical depth based on model HI column density and temperature. All model results are for volume filling factor unity ( $f_V = 1$ ) and cloud thickness 40 kpc. Yellow vertical line indicates radii within which QSO masking could impact results.

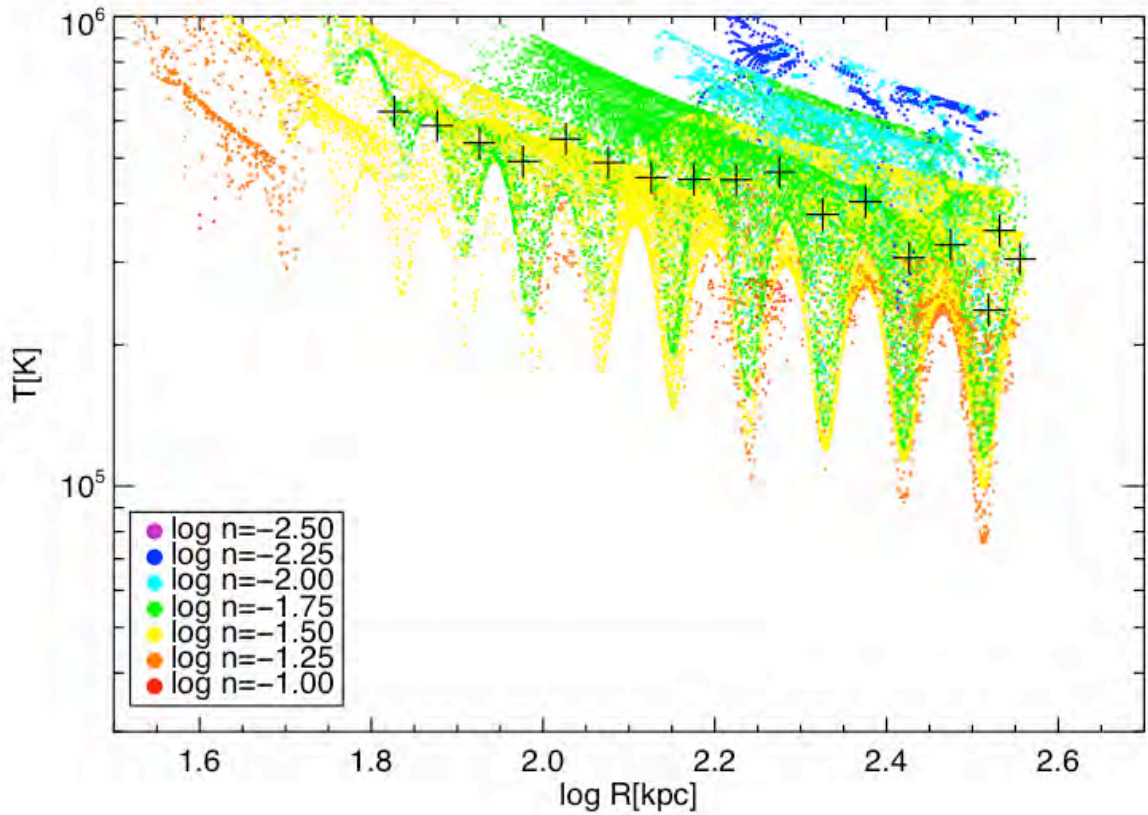

**Figure 26. Temperature vs. radial distance for complete set of cloud densities, CLOUDY model.** Colors indicate density. All model results are for volume filling factor unity ( $f_V = 1$ ) and cloud thickness 40 kpc. Cross symbols indicate mean temperatures inferred for the data vs. radius. The temperatures are high in the CQM, compared to typical photoionized gas, because the QSO is hyperluminous, and the ionization parameter U>10.

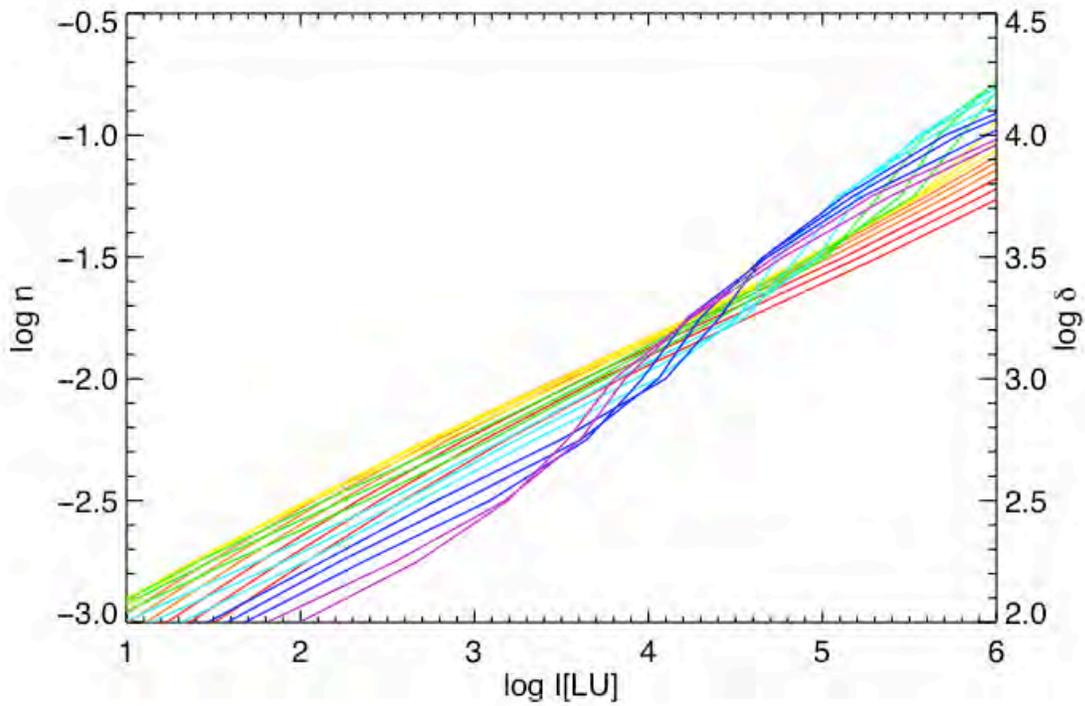

**Figure 27. Inferred density given surface brightness from CLOUDY model.** Line show for the intrinsic cloud density vs. the observed intensity for each radial position from the QSO. Radii range from 30 kpc to 600 kpc (red to violet) in equal logarithmic steps. Without knowing the physical radial distance, the error in the conversion from intensity to density is ~0.3 dex. Right axis shows the overdensity parameter for standard baryon fraction at this redshift. All curves assume unit volume filling factor ($f_V = 1$) and cloud thickness 40 kpc.

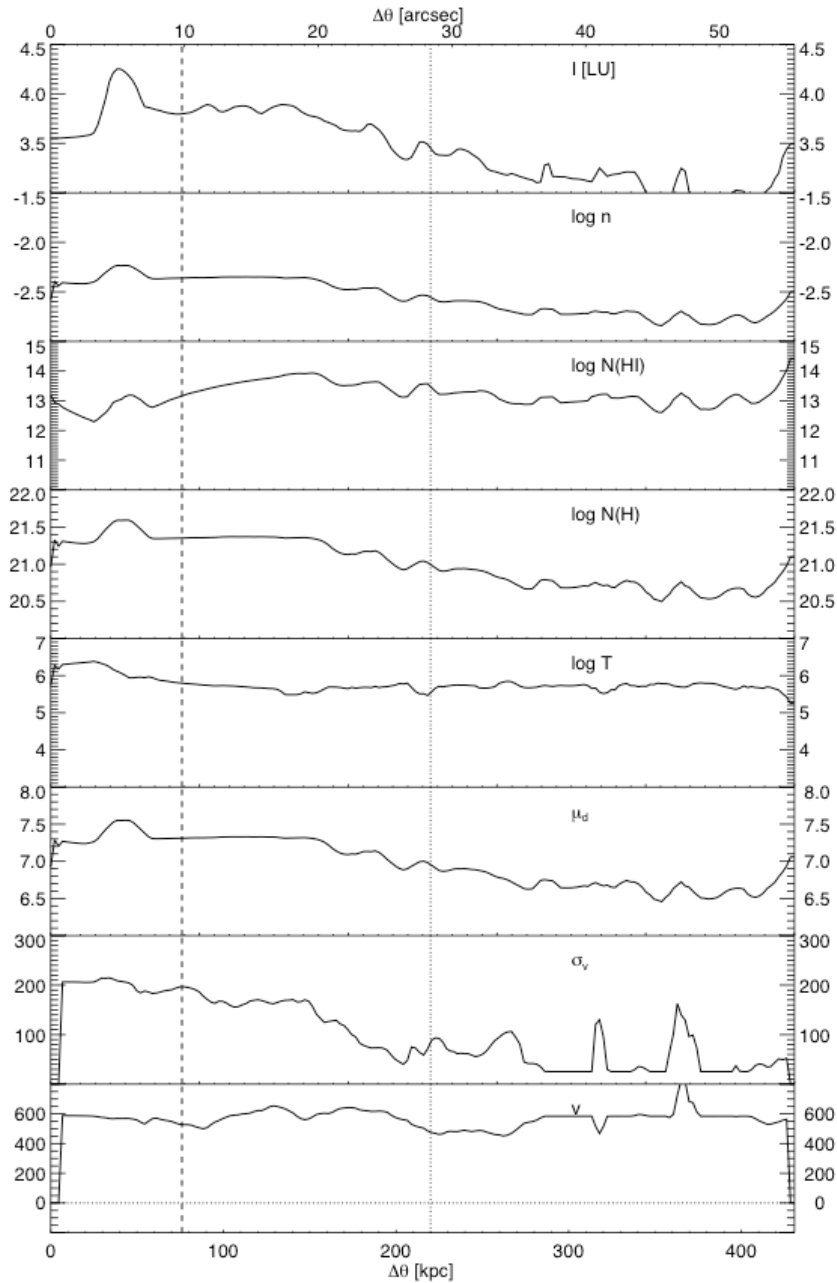

**Figure 28. Inferred physical properties along filament 1.** All curves assume unit volume filling factor ($f_V = 1$) and cloud thickness 40 kpc. The vertical dotted line at $\Delta\theta = 220$ kpc corresponds to the virial radius based on the mass estimate. The vertical dashed line indicates the inner region near QSO that has been masked. The horizontal dotted line in the velocity (bottom) plot indicates the QSO systemic velocity.

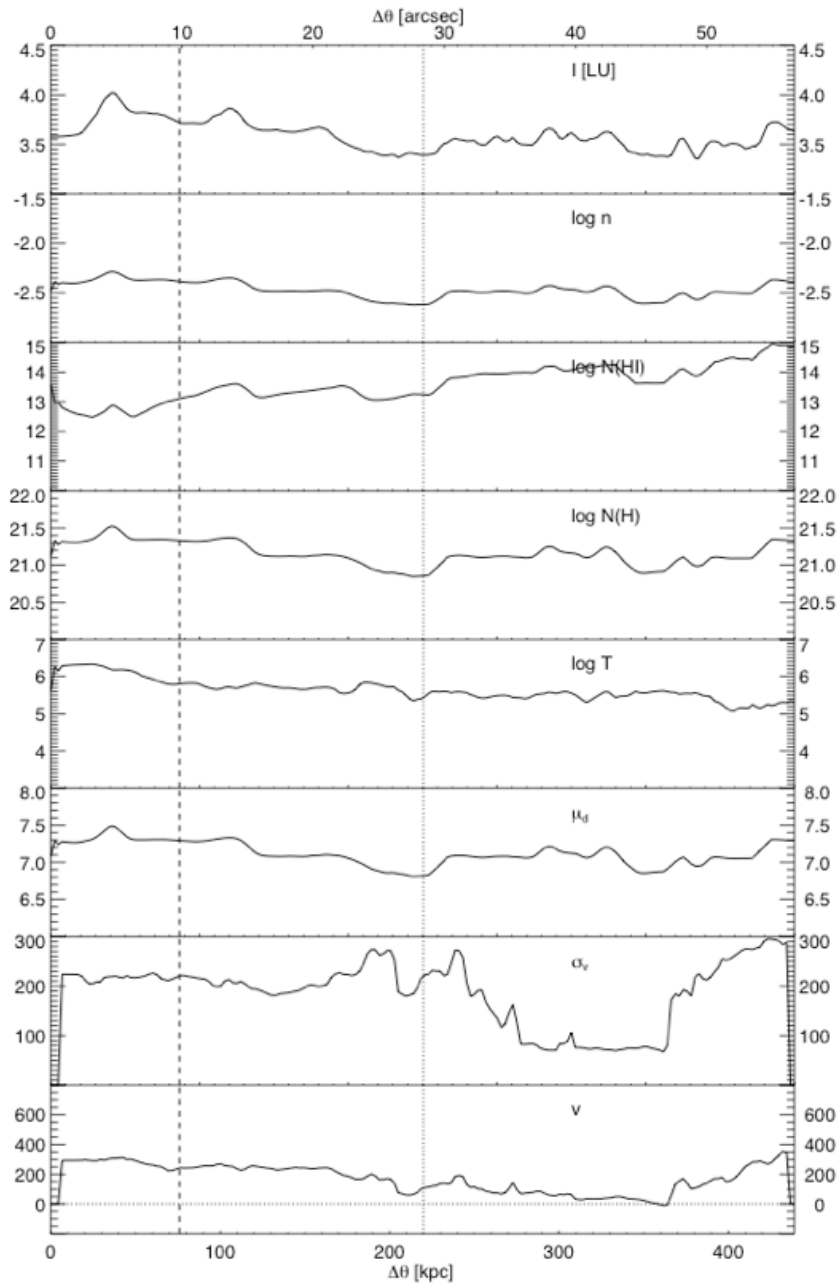

**Figure 29. Inferred physical properties along filament 2.** All curves assume unit volume filling factor ($f_V = 1$) and cloud thickness 40 kpc. The vertical dashed line indicates the inner region near QSO that has been masked. The horizontal dotted line in the velocity (bottom) plot indicates the QSO systemic velocity.

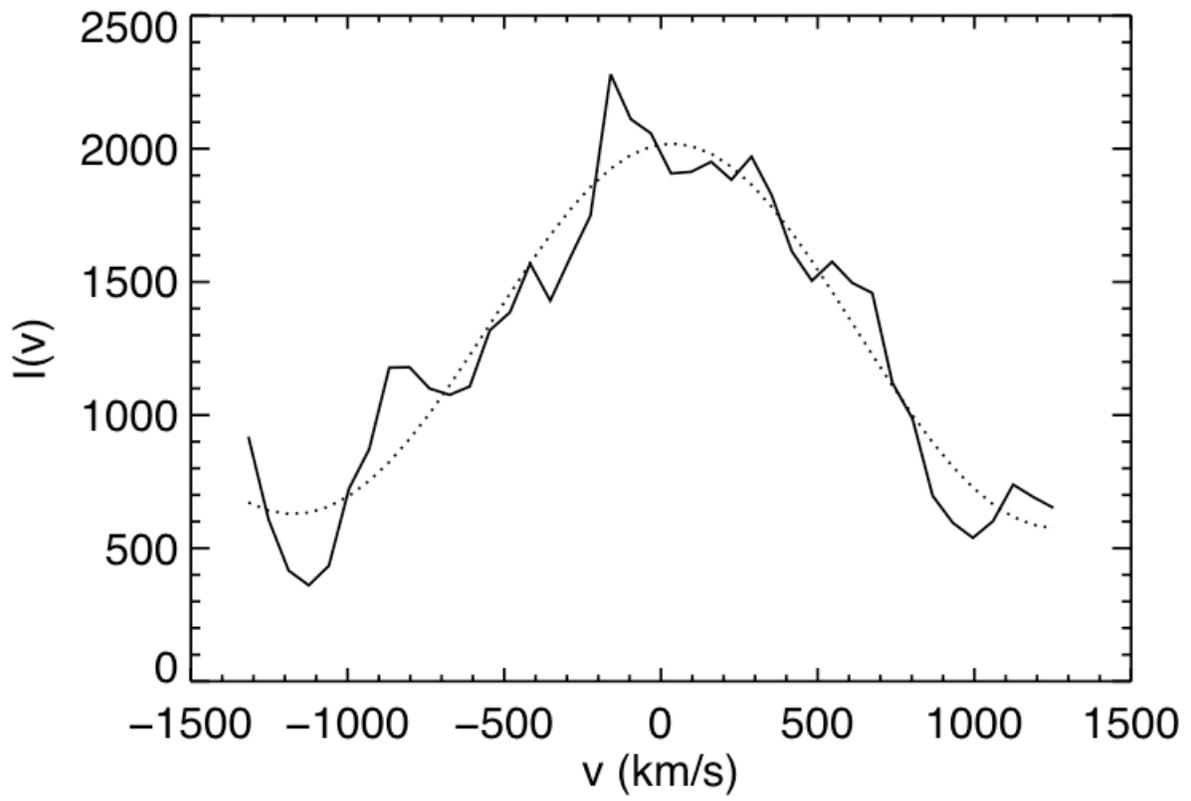

**Figure 30. Velocity profile of intensity, along with Gaussian fit showing 1D line-of-sight velocity dispersion σ$_v$=700 km/s.**

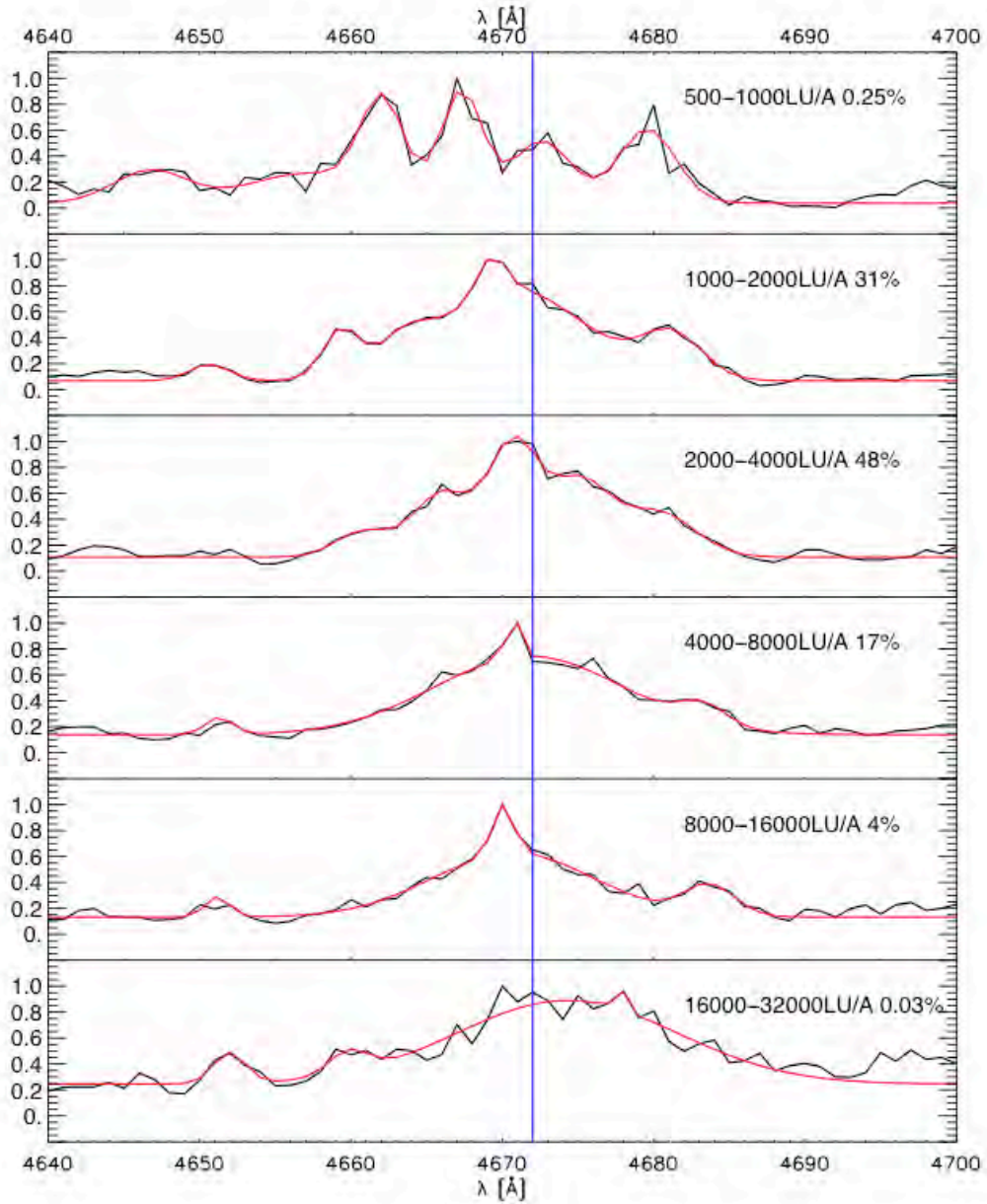

**Figure 31. Spectra of emission binned by intensity.** This figure shows the normalized spectrum binned by intensity (LU/Å) as shown in each panel, using the smoothed data cube. The spectra are fit with multiple Gaussian distributions, shown in red. Line centroid and velocity width from fits are used in the following figure. The fraction of the estimated total mass (see text) corresponding to each intensity bin is shown after the intensity range.

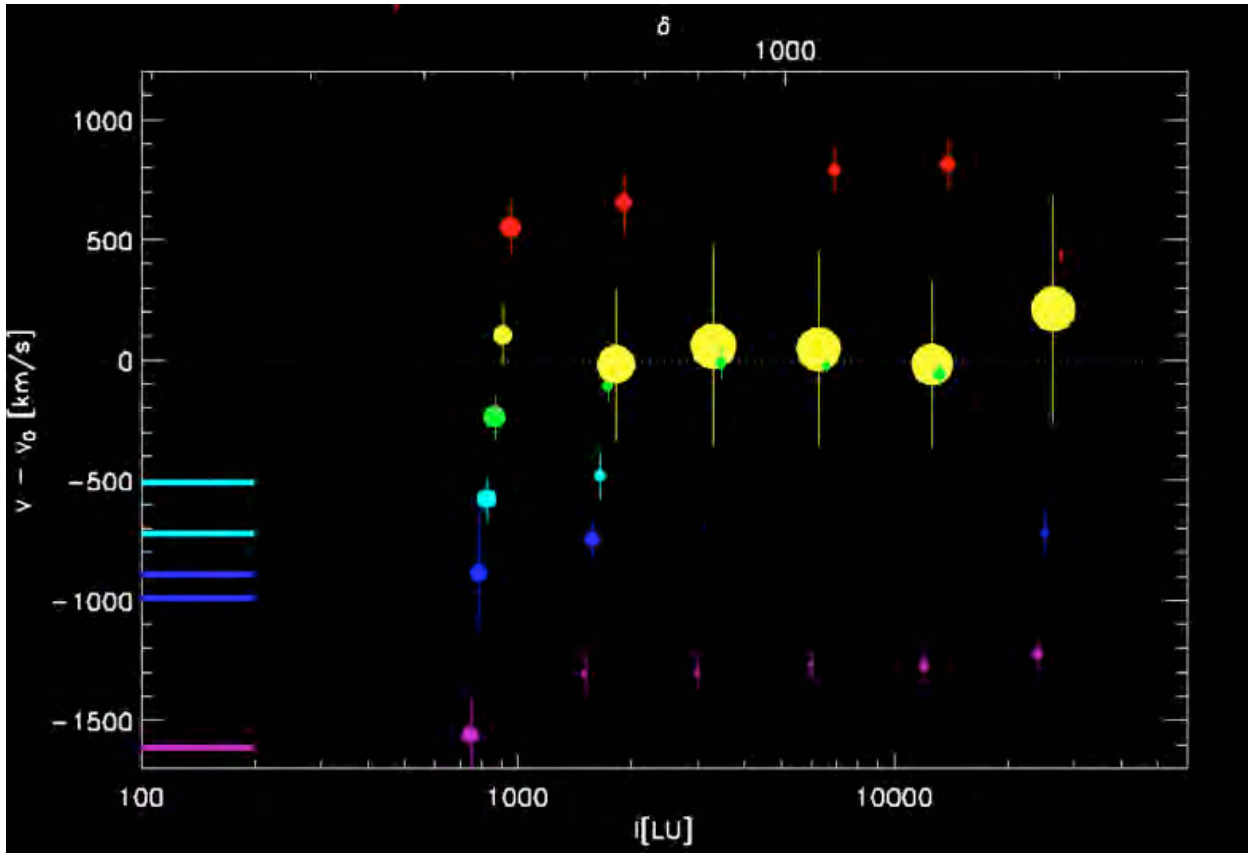

**Figure 32. Summary of kinematic components vs. intensity (density).** The spectral line fits to the intensity binned spectra shown in **Figure 31** are displayed in this plot vs. intensity (lower scale), and overdensity (upper scale, see **Figure 27**). Colors are used to separate distinct kinematic components. Circle location in velocity is best fit central velocity for each component. Vertical bar shows +/- 1-sigma velocity width of Gaussian fit. Size of dot indicates fraction of total emission in this component (from Gaussian height fit). Horizontal lines at left show QSO absorption line components.

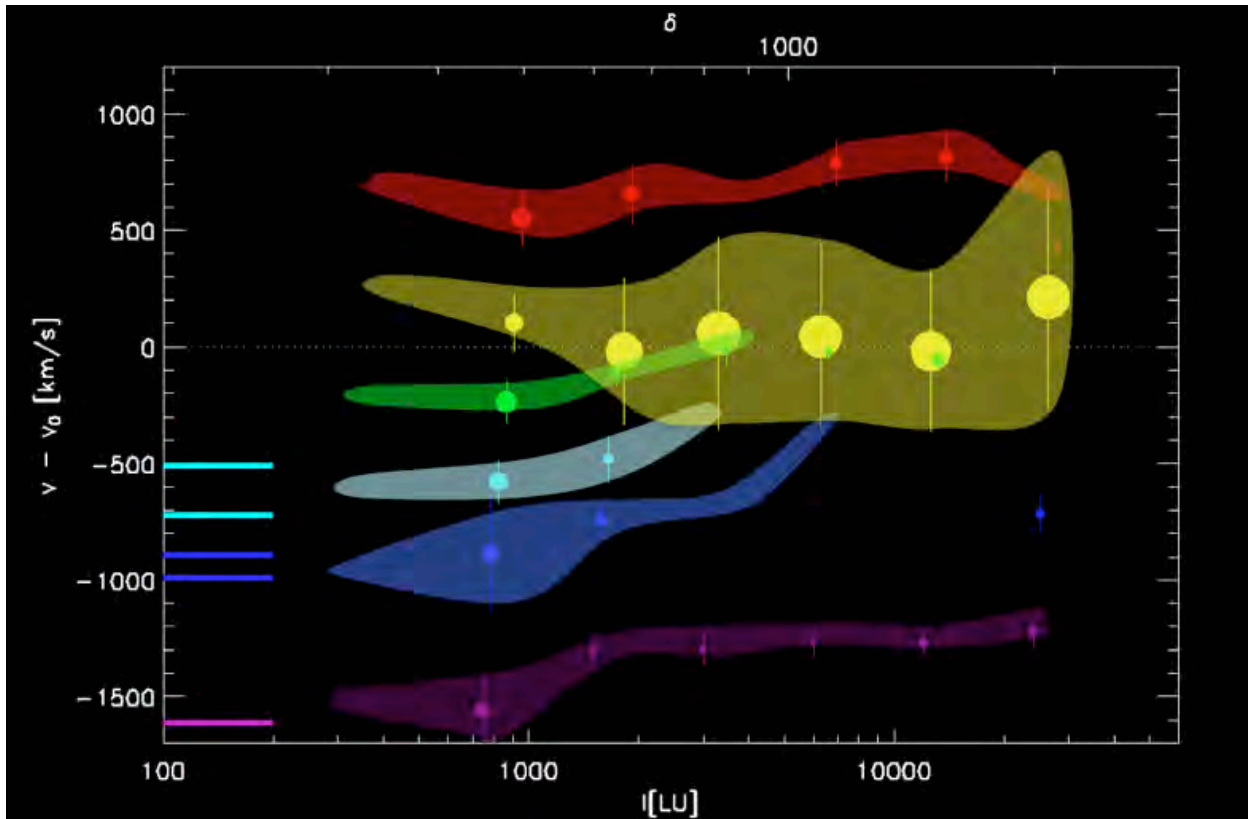

**Figure 33. Summary of kinematic components vs. intensity (density) with overlay.** The spectral line fits to the intensity binned spectra shown in **Figure 31** are displayed in this plot vs. intensity (lower scale), and overdensity (upper scale). Colors are used to separate distinct kinematic components. Circle location in velocity is best fit central velocity for each component. Vertical bar shows +/- 1-sigma velocity width of Gaussian fit. Size of dot indicates fraction of total emission in this component (from Gaussian height fit). Horizontal lines at left show QSO absorption line components. In this version of the figure we have overlaid a cartoon (shaded regions) suggesting that higher densities are kinematically evolved in the sense of harboring fewer and broader kinematic components. The implication is that the material at higher density is more virialized.